\documentclass[aps,prd,groupedaddress,showpacs,nofootinbib,twocolumn]{revtex4}
\usepackage{graphicx,amssymb}

\begin{document}

\title{Cosmography of $f(R)$ gravity}

\author{S. Capozziello\footnote{{\tt capozziello@na.infn.it}}, V.F. Cardone
\footnote{{\tt winnyenodrac@gmail.com}}, V. Salzano\footnote{{\tt
salzano@na.infn.it}}}

\affiliation{Dipartimento di Scienze Fisiche, Univ. di Napoli
"Federico II" and  INFN, Sez. di Napoli,   Compl. Univ. di Monte
S. Angelo, Ed. N, via Cinthia, 80126 - Napoli, Italy}
\date{\today}

\begin{abstract}

It is nowadays accepted that the universe is undergoing a phase of
accelerated expansion as tested by the Hubble diagram of Type Ia
Supernovae (SNeIa) and several LSS observations. Future SNeIa
surveys and other probes will make it possible to better
characterize the dynamical state of the universe renewing the
interest in cosmography which allows a model independent analysis
of the distance\,-\,redshift relation. On the other hand, fourth
order theories of gravity, also referred to as $f(R)$ gravity,
have attracted a lot of interest since they could be able to
explain the accelerated expansion without  any dark energy. We
show here how it is possible to relate the cosmographic parameters
(namely the deceleration $q_0$, the jerk $j_0$, the snap $s_0$ and
the lerk $l_0$ parameters) to the present day values of $f(R)$ and
its derivatives $f^{(n)}(R) = d^nf/dR^n$ (with $n = 1, 2, 3$) thus
offering a new tool to constrain such higher order models.  Our
analysis thus offers the possibility to relate the model
independent results coming from cosmography to the theoretically
motivated assumptions of $f(R)$ cosmology.

\end{abstract}


\maketitle

\section{Introduction}

As soon as astrophysicists realized that Type Ia Supernovae (SNeIa) were
standard candles, it appeared evident that their high luminosity should
make it possible to build a Hubble diagram, i.e. a plot of the
distance\,-\,redshift relation, over some cosmologically interesting
distance ranges. Motivated by this attractive consideration, two
independent teams started SNeIa surveys leading to the unexpected discovery
that the universe expansion is speeding up rather than decelerating
\cite{SNeIaFirst}. This surprising result has now been strengthened by more
recent data coming from SNeIa surveys
\cite{SNeIaSec,Riess04,R06,SNLS,ESSENCE,D07}, large scale structure
\cite{LSS} and cosmic microwave background (CMBR) anisotropy spectrum
\cite{CMBR,wmap,WMAP3}. This large dataset coherently points toward the
picture of a spatially flat universe undergoing an accelerated expansion
driven by a dominant negative pressure fluid, typically referred to as {\it
dark energy} \cite{copeland}.

While there is a wide consensus on the above scenario depicted by such good
quality data, there is a similarly wide range of contrasting proposals to
solve the dark energy puzzle. Surprisingly, the simplest explanation,
namely the cosmological constant $\Lambda$ \cite{Lambda}, is also the best
one from a statistical point of view \cite{LCDMtest}. Unfortunately, the
well known coincidence and 120 orders of magnitude problems render
$\Lambda$ a rather unattractive solution from a theoretical point of view.
Inspired by the analogy with inflation, a scalar field $\phi$, dubbed {\it
quintessence} \cite{QuintRev}, has then been proposed to give a dynamical
$\Lambda$ term in order to both fit the data and avoid the above problems.
However, such models are still plagued by difficulties on their own, such
as the almost complete freedom in the choice of the scalar field potential
and the fine tuning of the initial conditions. Needless to say, a plethora
of alternative models are now on the market all sharing the main property
to be in agreement with observations, but relying on completely different
physics.

Notwithstanding their differences, all the dark energy based theories
assume that the observed acceleration is the outcome of the action of an up
to now undetected ingredient to be added to the cosmic pie. In terms of the
Einstein equations, $G_{\mu \nu} = \chi T_{\mu \nu}$, such models are
simply modifying the right hand side including in the stress\,-\,energy
tensor something more than the usual matter and radiation components.

As a radically different approach, one can also try to leave unchanged the
source side, but rather modifying the left hand side. In a sense, one is
therefore interpreting cosmic speed up as a first signal of the breakdown
of the laws of physics as described by the standard General Relativity
(GR). Since this theory has been experimentally tested only up to the Solar
System scale, there is no a priori theoretical motivation to extend its
validity to extraordinarily larger scales such as the cosmological ones
(e.g. the last scattering surface!). Extending GR, not giving up to its
positive results, opens the way to a large class of alternative theories of
gravity ranging from extra\,-\,dimensions \cite{DGP} to nonminimally
coupled scalar fields \cite{stbook,stnoi}. In particular, we will be
interested here in fourth order theories \cite{fognoi,fogaltri} based on
replacing the scalar curvature $R$ in the Hilbert--Einstein action with a
generic analytic function $f(R)$ which should be reconstructed starting
from data and physically motivated issues. Also referred to as $f(R)$
gravity, these models have been shown to be able to both fit the
cosmological data and evade the Solar System constraints in several
physically interesting cases \cite{Hu,Starobinsky,Appleby,Odintsov,Tsuji}.

It is worth noting that both dark energy models and modified gravity
theories have shown to be in agreement with the data. As a consequence,
unless higher precision probes of the expansion rate and the growth of
structure will be available, these two rival approaches could not be
discriminated. This confusion about the theoretical background suggests
that a more conservative approach to the problem of cosmic acceleration,
relying on as less model dependent quantities as possible, is welcome.  A
possible solution could be to come back to the cosmography \cite{W72}
rather than finding out solutions of the Friedmann equations and testing
them. Being only related to the derivatives of the scale factor, the
cosmographic parameters make it possible to fit the data on the
distance\,-\,redshift relation without any {\it a priori} assumption on the
underlying cosmological model: in this case, the only assumption is that
the metric is the Robertson\,-\,Walker one (and hence not relying on the
solution of cosmic equations). Almost a century after Hubble discovery of
the expansion of the universe, we could  now  extend cosmography beyond the
search for the value of the Hubble constant. The SNeIa Hubble diagram
extends up to $z = 1.7$ thus invoking the need for, at least, a fifth order
Taylor expansion of the scale factor in order to give a reliable
approximation of the distance\,-\,redshift relation. As a consequence, it
could be, in principle, possible to estimate up to five cosmographic
parameters, although the still too small dataset available does not allow
to get a precise and realistic determination of all of them.

Once these quantities have been determined, one could use them to put
constraints on the models. In a sense, we are reverting the usual approach
consisting in deriving the cosmographic parameters as a sort of byproduct
of an assumed theory. Here, we follow the other way around expressing the
model characterizing quantities as a function of the cosmographic
parameters. Such a program is particularly suited for the study of fourth
order theories of gravity. As is well known, the mathematical difficulties
entering the solution of fourth order field equations make it quite
problematic to find out analytical expressions for the scale factor and
hence predict the values of the cosmographic parameters. A key role in
$f(R)$ gravity is played by the choice of the $f(R)$ function. Under quite
general hypotheses, we will derive useful relations among the cosmographic
parameters and the present day value of $f^{(n)}(R) = d^nf/dR^n$, with $n =
0, \ldots, 3$, whatever $f(R)$ is\footnote{As an important remark, we
stress that our derivation will rely on the metric formulation of $f(R)$
theories, while we refer the reader to \cite{poplawski} for a similar work
in the Palatini approach.}. Once the cosmographic parameters will be
determined, this method will allow us to investigate the cosmography of
$f(R)$ theories.

The layout of the paper is as follows. Sects. II and III are devoted to
introducing the basic notions of the cosmographic parameters and $f(R)$
gravity, respectively, summarizing the main formulae we will use later.
Sect. IV contains the main result of the paper demonstrating how the $f(R)$
derivatives can be related to the cosmographic parameters. Since these
latter are not well determined today, we will discuss, in Sect. V, how
these formulae can be adapted to a different parameterization relying on
expressing the cosmographic parameters in terms of a phenomenological
assumption for the dark energy equation of state. Sect. VI illustrates a
possible application of the relation among $f(R)$ derivatives and
cosmographic parameters showing how one can constrain the parameters of a
given $f(R)$ model. Since future data will likely determine with a
sufficient precision at least the first two cosmographic parameters, it is
worth estimating how this will impact on the determination of the $f(R)$
quantities, which is the argument of Sect. VII. We then summarize and
conclude in Sect. VIII.

\section{Cosmographic parameters}

Standard candles (such as SNeIa and, to a limited extent, gamma
ray bursts) are ideal tools in modern cosmology since they make it
possible to reconstruct the Hubble diagram, i.e. the
redshift\,-\,distance relation up to high redshift values. It is
then customary to assume a parameterized model (such as the
concordance $\Lambda$CDM one, or any other kind of dark energy
scenario) and contrasting it against the data to check its
viability and constraints its characterizing parameters. As it is
clear, such an approach is model dependent so that some doubts
always remain on the validity of the constraints on derived
quantities as the present day values of the deceleration parameter
and the age of the universe. In order to overcome such a problem,
one may resort to cosmography, i.e. expanding the scale factor in
Taylor series with respect to the cosmic time \cite{W72}. Such an
expansions leads to a distance\,-\,redshit relation which only
relies on the assumption of the Robertson\,-\,Walker metric thus
being fully model independent since it does not depend on the
particular form of the solution of cosmic equations. To this aim,
it is convenient to introduce the following functions
\cite{W72,Visser}\,:

\begin{equation}
\begin{array}{l}
\displaystyle{H = \frac{1}{a} \frac{da}{dt}} \\ ~ \\ \displaystyle{q = - \frac{1}{a} \frac{d^2a}{dt^2} \ H^{-2}}
\\ ~ \\ \displaystyle{j = \frac{1}{a} \frac{d^3a}{dt^3} \ H^{-3}} \\ ~ \\
\displaystyle{s = \frac{1}{a} \frac{d^4a}{dt^4} \ H^{-4}} \\ ~ \\
\displaystyle{l = \frac{1}{a} \frac{d^5a}{dt^5} \ H^{-5}}
\end{array}
\label{eq: cosmopar}
\end{equation}
which are usually referred to as the {\it Hubble, deceleration, jerk, snap}
and {\it lerk} parameters \cite{Dabro}, respectively\footnote{Note that the
use of the jerk parameter to discriminate between different models was also
proposed in \cite{SF} in the context of the {\it statefinder}
parametrization.}. Their present day values (which we will denote with a
subscript $0$) may be used to characterize the evolutionary status of the
Universe. For instance, $q_0 < 0$ denotes an accelerated expansion, while
$j_0$ allows to discriminate among different accelerating models.

It is then a matter of algebra to demonstrate the following useful
relations\,:

\begin{equation}
\dot{H} = -H^2 (1 + q) \ ,
\label{eq: hdot}
\end{equation}

\begin{equation}
\ddot{H} = H^3 (j + 3q + 2) \ ,
\label{eq: h2dot}
\end{equation}

\begin{equation}
\dddot{H} = H^4 \left [ s - 4j - 3q (q + 4) - 6 \right ] \ ,
\label{eq: h3dot}
\end{equation}

\begin{equation}
d^4H/dt^4 = H^5 \left [ l - 5s + 10 (q + 2) j + 30 (q + 2) q + 24 \right ]
\ ,
\label{eq: h4dot}
\end{equation}
where a dot denotes derivative with respect to the cosmic time $t$.
Eqs.(\ref{eq: hdot})\,-\,(\ref{eq: h4dot}) make it possible to relate the
derivative of the Hubble parameter to the other cosmographic parameters.
The distance\,-\,redshift relation may then be obtained starting from the
Taylor expansion of $a(t)$ along the lines described in
\cite{Visser,WM04,CV07}. The result for the fifth order is reported in
Appendix A.

It is worth stressing that the definition of the cosmographic parameters
only relies on the assumption of the Robertson\,-\,Walker metric. As such,
it is however difficult to state a priori to what extent the fifth order
expansion provides an accurate enough description of the quantities of
interest. Actually, the number of cosmographic parameters to be used
depends on the problem one is interested in. As we will see later, we are
here concerned only with the SNeIa Hubble diagram so that we have to check
that the distance modulus $\mu_{cp}(z)$ obtained using the fifth order
expansion of the scale factor is the same (within the errors) as the one
$\mu_{DE}(z)$ of the underlying physical model. Being such a model of
course unknown, one can adopt a phenomenological parameterization for the
dark energy\footnote{Note that one can always use a phenomenological dark
energy model to get a reliable estimate of the scale factor evolution even
if the correct model is a fourth order one.} EoS and look at the percentage
deviation $\Delta \mu/\mu_{DE}$ as function of the EoS parameters. We have
carried out such exercise using the CPL model introduced later and verified
that $\Delta \mu/\mu_{DE}$ is an increasing function of $z$ (as expected),
but still remains smaller than $2\%$ up to $z \sim 2$ over a wide range of
the CPL parameter space. On the other hand, halting the Taylor expansion to
a lower order may introduce significant deviation for $z > 1$ that can
potentially bias the analysis if the measurement errors are as small as
those predicted for future SNeIa surveys. We are therefore confident that
our fifth order expansion is both sufficient to get an accurate distance
modulus over the redshift range probed by SNeIa and necessary to avoid
dangerous biases.

\section{$f(R)$ gravity}

Much interest has been recently devoted to a form of quintessence induced
by curvature  according to which the present universe is filled by
pressureless dust matter only and the acceleration is the result of the
modified Friedmann equations obtained by replacing the Ricci  curvature
scalar $R$ with a generic function $f(R)$ in the gravity action
\cite{fognoi,fogaltri}. Under the assumption of a flat universe, the Hubble
parameter is therefore determined by\footnote{We use here natural units
such that $8
\pi G = 1$.}\,:

\begin{equation}
H^2 = \frac{1}{3} \left [ \frac{\rho_m}{f'(R)} + \rho_{curv} \right ]
\label{eq: hfr}
\end{equation}
where the prime denotes derivative with respect to $R$ and $\rho_{curv}$ is
the energy density of an {\it effective curvature fluid}\footnote{Note that
the name {\it curvature fluid} does not refer to the FRW curvature
parameter $k$, but only takes into account that such a term is a
geometrical one related to the scalar curvature $R$.}\,:

\begin{equation}
\rho_{curv} = \frac{1}{f'(R)} \left \{ \frac{1}{2} \left [ f(R)  - R f'(R) \right ]
- 3 H \dot{R} f''(R) \right \} \ .
\label{eq: rhocurv}
\end{equation}
Assuming there is no interaction between the matter and the
curvature terms (we are in the so-called {\it Jordan frame}), the
matter continuity equation gives the usual scaling $\rho_M =
\rho_M(t = t_0) a^{-3} = 3 H_0^2 \Omega_M a^{-3}$, with $\Omega_M$
the present day matter density parameter. The continuity equation
for $\rho_{curv}$ then reads\,:

\begin{equation}
\dot{\rho}_{curv} + 3 H (1 + w_{curv}) \rho_{curv}  =
\frac{3 H_0^2 \Omega_M \dot{R} f''(R)}{\left [ f'(R) \right ]^2}  a^{-3}
\label{eq: curvcons}
\end{equation}
with

\begin{equation}
w_{curv} = -1 + \frac{\ddot{R} f''(R) + \dot{R} \left [ \dot{R} f'''(R) - H
f''(R) \right ]} {\left [ f(R) - R f'(R) \right ]/2 - 3 H \dot{R} f''(R)}
\label{eq: wcurv}
\end{equation}
the barotropic factor of the curvature fluid. It is worth noticing
that the curvature fluid quantities $\rho_{curv}$ and $w_{curv}$
only depends on $f(R)$ and its derivatives up to the third order.
As a consequence, considering only their present day values (which
may be naively obtained by replacing $R$ with $R_0$ everywhere),
two $f(R)$ theories sharing the same values of $f(R_0)$,
$f'(R_0)$, $f''(R_0)$, $f'''(R_0)$ will be degenerate from this
point of view\footnote{One can argue that this is not strictly
true since different $f(R)$ theories will lead to different
expansion rate $H(t)$ and hence different present day values of
$R$ and its derivatives. However, it is likely that two $f(R)$
functions that exactly match each other up to the third order
derivative today will give rise to the same $H(t)$ at least for $t
\simeq t_0$ so that $(R_0, \dot{R}_0, \ddot{R}_0)$ will be almost
the same.}.

Combining Eq.(\ref{eq: curvcons}) with Eq.(\ref{eq: hfr}), one finally gets
the following {\it master equation} for the Hubble parameter\,:

\begin{eqnarray}
\dot{H} & = & -\frac{1}{2 f'(R)} \left \{ 3 H_0^2 \Omega_M a^{-3} + \ddot{R} f''(R)+ \right . \nonumber \\
~ & ~ & \left . + \dot{R} \left [ \dot{R} f'''(R) - H f''(R) \right ] \right \} \ .
\label{eq: presingleeq}
\end{eqnarray}
Expressing the scalar curvature $R$ as function of the Hubble parameter
as\,:

\begin{equation}
R = - 6 \left ( \dot{H} + 2 H^2 \right )
\label{eq: rvsh}
\end{equation}
and inserting the result into Eq.(\ref{eq: presingleeq}), one ends
with a fourth order nonlinear differential equation for the scale
factor $a(t)$ that cannot be easily solved also for the simplest
cases (for instance, $f(R) \propto R^n$). Moreover, although
technically feasible, a numerical solution of Eq.(\ref{eq:
presingleeq}) is plagued by the large uncertainties on the
boundary conditions (i.e., the present day values of the scale
factor and its derivatives up to the third order) that have to be
set to find out the scale factor.

\section{$f(R)$ derivatives vs cosmography}

Motivated by these difficulties, we approach now the problem from a
different viewpoint. Rather than choosing a parameterized expression for
$f(R)$ and then numerically solving Eq.(\ref{eq: presingleeq}) for given
values of the boundary conditions, we try to relate the present day values
of its derivatives to the cosmographic parameters $(q_0, j_0, s_0, l_0)$ so
that constraining them in a model independent way gives us a hint for what
kind of $f(R)$ theory could be able to fit the observed Hubble
diagram\footnote{Note that a similar analysis, but in the context of the
energy conditions in $f(R)$, has yet been presented in \cite{Bergliaffa}.
However, in that paper, the author give an expression for $f(R)$ and then
compute the snap parameter to be compared to the observed one. On the
contrary, our analysis does not depend on any assumed functional expression
for $f(R)$.}.

As a preliminary step, it is worth considering again the constraint
equation (\ref{eq: rvsh}). Differentiating with respect to $t$, we easily
get the following relations\,:

\begin{equation}
\begin{array}{l}
\displaystyle{\dot{R} = -6 \left ( \ddot{H} + 4 H \dot{H} \right )} \\ ~ \\
\displaystyle{\ddot{R} = -6 \left ( \dddot{H} + 4 H \ddot{H} + 4 \dot{H}^2 \right )} \\  ~ \\
\displaystyle{\dddot{R} = -6 \left ( d^4H/dt^4 + 4 H \dddot{H} + 12 \dot{H} \ddot{H} \right )} \\
\end{array}
\ .
\label{eq: prederr}
\end{equation}
Evaluating these at the present time and using Eqs.(\ref{eq:
hdot})\,-\,(\ref{eq: h4dot}), one finally gets\,:

\begin{equation}
R_0 = -6 H_0^2 (1 - q_0) \ ,
\label{eq: rz}
\end{equation}

\begin{equation}
\dot{R}_0 = -6 H_0^3 (j_0 - q_0 - 2) \ ,
\label{eq: rdotz}
\end{equation}

\begin{equation}
\ddot{R}_0 = -6 H_0^4 \left ( s_0 + q_0^2 + 8 q_0 + 6 \right ) \ ,
\label{eq: r2dotz}
\end{equation}

\begin{equation}
\dddot{R}_0 = -6 H_0^5 \left [ l_0 - s_0 + 2 (q_0 + 4) j_0 - 6 (3q_0 + 8) q_0 - 24 \right ] \ ,
\label{eq: r3dotz}
\end{equation}
which will turn out to be useful in the following.

Let us now come back to the expansion rate and master equations (\ref{eq:
hfr}) and (\ref{eq: presingleeq}). Since they have to hold along the full
evolutionary history of the universe, they naively hold also at the present
day. As a consequence, we may evaluate them in $t = t_0$ thus easily
obtaining\,:

\begin{eqnarray}
H_0^2 & = & \frac{H_0^2 \Omega_M}{f'(R_0)}   \nonumber \\ ~ & + &
\frac{f(R_0) - R_0 f'(R_0) - 6 H_0 \dot{R}_0 f''(R_0)}{6 f'(R_0)} \ ,
\label{eq: hfrz}
\end{eqnarray}

\begin{eqnarray}
- \dot{H}_0 & = & \frac{3 H_0^2 \Omega_M}{2 f'(R_0)} \nonumber \\ ~ & + & \frac{\dot{R}_0^2
f'''(R_0) + \left ( \ddot{R}_0 - H_0 \dot{R_0} \right ) f''(R_0)}{2
f'(R_0)} \ .
\label{eq: hdotfrz}
\end{eqnarray}
Using Eqs.(\ref{eq: hdot})\,-\,(\ref{eq: h4dot}) and (\ref{eq:
rz})\,-\,(\ref{eq: r3dotz}), we can rearrange Eqs.(\ref{eq: hfrz})
and (\ref{eq: hdotfrz}) as two relations among the Hubble constant
$H_0$ and the cosmographic parameters $(q_0, j_0, s_0)$, on one
hand, and the present day values of $f(R)$ and its derivatives up
to third order. However, two further relations are needed in order
to close the system and determine the four unknown quantities
$f(R_0)$, $f'(R_0)$, $f''(R_0)$, $f'''(R_0)$. A first one may be
easily obtained by noting that, inserting back the physical units,
the rate expansion equation reads\,:

\begin{displaymath}
H^2 = \frac{8 \pi G}{3 f'(R)} \left [\rho_m + \rho_{curv} f'(R) \right ]
\end{displaymath}
which clearly shows that, in $f(R)$ gravity, the Newtonian
gravitational constant $G$ is replaced by an effective (time
dependent) $G_{eff} = G/f'(R)$. On the other hand, it is
reasonable to assume that the present day value of $G_{eff}$ is
the same as the Newtonian one so that we get the simple
constraint\,:

\begin{equation}
G_{eff}(z = 0) = G \rightarrow f'(R_0) = 1 \ .
\label{eq: fpz}
\end{equation}
In order to get the fourth relation we need to close the system,
we first differentiate both sides of Eq.(\ref{eq: presingleeq})
with respect to $t$. We thus get\,:

\begin{eqnarray}
\ddot{H} & = & \frac{\dot{R}^2 f'''(R) + \left ( \ddot{R} - H \dot{R} \right ) f''(R)
+ 3 H_0^2 \Omega_M a^{-3}}{2 \left [ \dot{R} f''(R) \right ]^{-1} \left [
f'(R) \right ]^2} \nonumber \\ ~ & - & \frac{\dot{R}^3 f^{(iv)}(R) + \left
( 3 \dot{R} \ddot{R} - H \dot{R}^2 \right ) f'''(R)}{2 f'(R)} \nonumber \\
~ & - & \frac{\left ( \dddot{R} - H \ddot{R} + \dot{H} \dot{R} \right )
f''(R) - 9 H_0^2 \Omega_M H a^{-3}}{2 f'(R)} \ ,
\label{eq: h2dotfr}
\end{eqnarray}
with $f^{(iv)}(R) = d^4f/dR^4$. Let us now suppose that $f(R)$ may be well
approximated by its third order Taylor expansion in $R - R_0$, i.e. we
set\,:

\begin{eqnarray}
f(R) & = & f(R_0) + f'(R_0) (R - R_0) +  \frac{1}{2} f''(R_0) (R - R_0)^2
\nonumber \\ ~ & + & \frac{1}{6} f'''(R_0) (R - R_0)^3 \ .
\label{eq: frtaylor}
\end{eqnarray}
In such an approximation, it is $f^{(n)}(R) = d^nf/R^n = 0$ for $n \ge 4$
so that naively $f^{(iv)}(R_0) = 0$. Evaluating then Eq.(\ref{eq: h2dotfr})
at the present day, we get\,:

\begin{eqnarray}
\ddot{H}_0 & = & \frac{\dot{R}_0^2 f'''(R_0) + \left ( \ddot{R}_0 - H_0 \dot{R}_0
\right ) f''(R_0) + 3 H_0^2 \Omega_M}{2 \left [ \dot{R}_0 f''(R_0) \right ]^{-1} \left [
f'(R_0) \right ]^2} \nonumber \\ ~ & - & \frac{ \left ( 3 \dot{R}_0
\ddot{R}_0 - H \dot{R}_0^2 \right ) f'''(R_0)}{2 f'(R_0)}
\nonumber \\ ~ & - & \frac{\left ( \dddot{R}_0 - H_0 \ddot{R}_0 + \dot{H}_0 \dot{R}_0
\right ) f''(R_0) - 9 H_0^3 \Omega_M}{2 f'(R_0)} \ .
\label{eq: h2dotfrz}
\end{eqnarray}
We can now schematically proceed as follows. Evaluate
Eqs.(\ref{eq: hdot})\,-\,(\ref{eq: h4dot}) at $z = 0$ and plug
these relations into the left hand sides of Eqs.(\ref{eq: hfrz}),
(\ref{eq: hdotfrz}), (\ref{eq: h2dotfrz}). Insert Eqs.(\ref{eq:
rz})\,-\,(\ref{eq: r3dotz}) into the right hand sides of these
same equations so that only the cosmographic parameters $(q_0,
j_0, s_0, l_0)$ and the $f(R)$ related quantities enter both sides
of these relations. Finally, solve them under the constraint
(\ref{eq: fpz}) with respect to the present day values of $f(R)$
and its derivatives up to the third order. After some  algebra,
one ends up with the desired result\,:

\begin{equation}
\frac{f(R_0)}{6 H_0^2} = - \frac{{\cal{P}}_0(q_0, j_0, s_0, l_0) \Omega_M +
{\cal{Q}}_0(q_0, j_0, s_0, l_0)}{{\cal{R}}(q_0, j_0, s_0, l_0)} \ ,
\label{eq: f0z}
\end{equation}

\begin{equation}
f'(R_0) = 1 \ ,
\label{eq: f1z}
\end{equation}

\begin{equation}
\frac{f''(R_0)}{\left ( 6 H_0^2 \right )^{-1}} = - \frac{{\cal{P}}_2(q_0,
j_0, s_0) \Omega_M + {\cal{Q}}_2(q_0, j_0, s_0)}{{\cal{R}}(q_0, j_0, s_0,
l_0)} \ ,
\label{eq: f2z}
\end{equation}

\begin{equation}
\frac{f'''(R_0)}{\left ( 6 H_0^2 \right )^{-2}} = - \frac{{\cal{P}}_3(q_0,
j_0, s_0, l_0) \Omega_M + {\cal{Q}}_3(q_0, j_0, s_0, l_0)}{(j_0 - q_0 - 2)
{\cal{R}}(q_0, j_0, s_0, l_0)} \ ,
\label{eq: f3z}
\end{equation}
where we have defined\,:

\begin{eqnarray}
{\cal{P}}_0 & = & (j_0 - q_0 - 2) l_0 \nonumber \\ ~ & - & (3s_0 + 7j_0 +
6q_0^2 + 41q_0 + 22) s_0 \nonumber \\ ~ & - & \left [ (3q_0 + 16) j_0 +
20q_0^2 + 64q_0 + 12 \right ] j_0 \nonumber \\ ~ &  - & \left ( 3q_0^4 +
25q_0^3 + 96q_0^2 + 72q_0 + 20 \right ) \ ,
\label{eq: defp0}
\end{eqnarray}

\begin{eqnarray}
{\cal{Q}}_0 & = & (q_0^2 - j_0 q_0 + 2q_0) l_0 \nonumber \\ ~ & + &
\left [ 3q_0s_0 + (4q_0 + 6) j_0 + 6q_0^3 + 44q_0^2 + 22q_0 - 12 \right ] s_0 \nonumber \\
~ & + & \left [ 2j_0^2 + (3q_0^2 + 10q_0 - 6) j_0 + 17q_0^3 + 52q_0^2 +
54q_0 \right . \nonumber  \\ ~ & + & \ \left . 36 \right ] j_0 + 3q_0^5 +
28q_0^4 + 118q_0^3 + 72q_0^2 - 76q_0 \nonumber \\ ~ & - & 64 \ ,
\label{eq: defq0}
\end{eqnarray}

\begin{equation}
{\cal{P}}_2 = 9 s_0 + 6 j_0 + 9q_0^2 + 66q_0 + 42 \ ,
\label{eq: defp2}
\end{equation}

\begin{eqnarray}
{\cal{Q}}_2 & = & - \left \{ 6 (q_0 + 1) s_0 \right . \nonumber \\ ~ & + &
\left [ 2j_0 - 2 (1 - q_0) \right ] j_0 \nonumber \\ ~ & + & \left . 6q_0^3 + 50q_0^2
+ 74q_0 + 32 \right \} \ ,
\label{eq: defq2}
\end{eqnarray}

\begin{equation}
{\cal{P}}_3 = 3 l_0  + 3 s_0 - 9(q_0 + 4) j_0 - (45q_0^2 + 78q_0 + 12) \ ,
\label{eq: defp3}
\end{equation}

\begin{eqnarray}
{\cal{Q}}_3 & = & - \left \{ 2 (1 + q_0) l_0 \right . \nonumber \\ ~ & + &
2 (j_0 + q_0) s_0 \nonumber \\ ~ & - & \left ( 2j_0 + 4q_0^2 + 12q_0 + 6
\right ) j_0 \nonumber \\ ~ & - & \left . (30q_0^3 + 84q_0^2 + 78q_0 + 24) \right \} \ ,
\label{eq: defq3}
\end{eqnarray}

\begin{eqnarray}
{\cal{R}} & = &  (j_0 - q_0 - 2) l_0 \nonumber \\ ~ & - & (3s_0 - 2j_0 +
6q_0^2 + 50q_0 + 40) s_0 \nonumber \\ ~ & + & \left [ (3q_0 + 10) j_0 +
11q_0^2 + 4q_0 -18 \right ] j_0 \nonumber \\ ~ & - & (3q_0^4 + 34q_0^3 +
246q_0 + 104) \ .
\label{eq: defr}
\end{eqnarray}
Eqs.(\ref{eq: f0z})\,-\,(\ref{eq: defr}) make it possible to
estimate the present day values of $f(R)$ and its first three
derivatives as function of the Hubble constant $H_0$ and the
cosmographic parameters $(q_0, j_0, s_0, l_0)$ provided a value
for the matter density parameter $\Omega_M$ is given. This is a
somewhat problematic point. Indeed, while the cosmographic
parameters may be estimated in a model independent way, the
fiducial value for $\Omega_M$ is usually the outcome of fitting a
given dataset in the framework of an assumed dark energy scenario.
However, it is worth noting that different models all converge
towards the concordance value $\Omega_M \simeq 0.25$ which is also
in agreement with astrophysical (model independent) estimates from
the gas mass fraction in galaxy clusters. On the other hand, it
has been proposed that $f(R)$ theories may avoid the need for dark
matter in galaxies and galaxy clusters \cite{fogdm}. In such a
case, the total matter content of the universe is essentially
equal to the baryonic one. According to the primordial elements
abundance and the standard BBN scenario, we therefore get
$\Omega_M \simeq \omega_b/h^2$ with $\omega_b = \Omega_b h^2
\simeq 0.0214$ \cite{Kirk} and $h$ the Hubble constant in units of
$100 {\rm km/s/Mpc}$. Setting $h = 0.72$ in agreement with the results of
the HST Key project \cite{Freedman}, we thus get $\Omega_M = 0.041$ for a
baryons only universe. We will therefore consider in the following both
cases when numerical estimates are needed.

It is worth noticing that $H_0$ only plays the role of a scaling
parameter giving the correct physical dimensions to $f(R)$ and its
derivatives. As such, it is not surprising that we need four
cosmographic parameters, namely $(q_0, j_0, s_0, l_0)$, to fix the
four $f(R)$ related quantities $f(R_0)$, $f'(R_0)$, $f''(R_0)$,
$f'''(R_0)$. It is also worth stressing that Eqs.(\ref{eq:
f0z})\,-\,({\ref{eq: f3z}) are linear in the $f(R)$ quantities so
that $(q_0, j_0, s_0, l_0)$ uniquely determine the former ones. On
the contrary, inverting them to get the cosmographic parameters as
function of the $f(R)$ ones, we do not get linear relations.
Indeed, the field equations in $f(R)$ theories are nonlinear
fourth order differential equations in the scale factor $a(t)$ so
that fixing the derivatives of $f(R)$ up to third order makes it
possible to find out a class of solutions, not a single one. Each
one of these solutions will be characterized by a different set of
cosmographic parameters thus explaining why the inversion of
Eqs.(\ref{eq: f0z})\,-\,(\ref{eq: defr}) does not give a unique
result for $(q_0, j_0, s_0, l_0)$.

As a final comment, we reconsider the underlying assumptions
leading to the above derived relations. While Eqs.(\ref{eq: hfrz})
and (\ref{eq: hdotfrz}) are exact relations deriving from a
rigorous application of the field equations, Eq.(\ref{eq:
h2dotfrz}) heavily relies on having approximated $f(R)$ with its
third order Taylor expansion (\ref{eq: frtaylor}). If this
assumption fails, the system should not be closed since a fifth
unknown parameter enters the game, namely $f^{(iv)}(R_0)$.
Actually, replacing $f(R)$ with its Taylor expansion is not
possible for all class of $f(R)$ theories. As such, the above
results only hold in those cases where such an expansion is
possible. Moreover, by truncating the expansion to the third
order, we are implicitly assuming that higher order terms are
negligible over the redshift range probed by the data. That is to
say, we are assuming that\,:

\begin{equation}
f^{(n)}(R_0) (R - R_0)^n << \sum_{m = 0}^{3}{\frac{f^{(m)}(R_0)}{m !} (R
- R_0)^m} \ \ {\rm for} \ n \ge 4
\label{eq: checkcond}
\end{equation}
over the redshift range probed by the data. Checking the validity of this
assumption is not possible without explicitly solving the field equations,
but we can guess an order of magnitude estimate considering that, for all
viable models, the background dynamics should not differ too much from the
$\Lambda$CDM one at least up to $z \simeq 2$. Using then the expression of
$H(z)$ for the $\Lambda$CDM model, it is easily to see that $R/R_0$ is a
quickly increasing function of the redshift so that, in order Eq.(\ref{eq:
checkcond}) holds, we have to assume that $f^{(n)}(R_0) << f'''(R_0)$ for
$n \ge 4$. This condition is easier to check for many analytical $f(R)$
models.

Once such a relation is verified, we have still to worry about
Eq.(\ref{eq: fpz}) relying on the assumption that the {\it
cosmological} gravitational constant is {\it exactly} the same as
the {\it local} one, i.e. the same as the one measured in the
laboratory and entering the Newtonian Poisson equation. Actually,
the {\it cosmological} gravitational constant should be identified
with the one entering the perturbation equations for a given
$f(R)$ model. Comparing the Newtonian $G_N$ and this {\it
cosmological} $G$, one could infer whether the $G$ entering the
background equations is the same as the local one. Although this
is outside our aims here, we can, in a first reasonable
approximation, argue that the condition $G_{local} = G_{cosmo}$
could be replaced by the weaker relation $G_{eff}(z = 0) = G (1 +
\varepsilon)$ with $\varepsilon << 1$. In this case, we should
repeat the derivation of Eqs.(\ref{eq: f0z})\,-\,(\ref{eq: f3z})
now using the condition $f'(R_0) = (1 + \varepsilon)^{-1}$. Taylor
expanding the results in $\varepsilon$ to the first order and
comparing with the above derived equations, we can estimate the
error induced by our assumption $\varepsilon = 0$. The resulting
expressions are too lengthy to be reported and depend in a
complicated way on the values of the matter density parameter
$\Omega_M$, the cosmographic parameters $(q_0, j_0, s_0, l_0)$ and
$\varepsilon$. However, we have numerically checked that the error
induced on $f(R_0)$, $f''(R_0)$, $f'''(R_0)$ are much lower than
$10\%$ for value of $\varepsilon$ as high as an unrealistic
$\varepsilon \sim 0.1$. We are therefore confident that our
results are reliable also under such conditions.

\section{$f(R)$ derivatives and CPL models}

A determination of $f(R)$ and its derivatives in terms of the
cosmographic parameters need for an estimate of these latter from
the data in a model independent way. Unfortunately, even in the
nowadays era of {\it precision cosmology}, such a program is still
too ambitious to give useful constraints on the $f(R)$
derivatives, as we will see later. On the other hand, the
cosmographic parameters may also be expressed in terms of the dark
energy density and EoS parameters so that we can work out what are
the present day values of $f(R)$ and its derivatives giving the
same $(q_0, j_0, s_0, l_0)$ of the given dark energy model. To
this aim, it is convenient to adopt a parameterized expression for
the dark energy EoS in order to reduce the dependence of the
results on any underlying theoretical scenario. Following the
prescription of the Dark Energy Task Force \cite{DETF}, we will
use the Chevallier\,-\,Polarski\,-\,Linder (CPL) parameterization
for the EoS setting \cite{CPL}\,:

\begin{equation}
w = w_0 + w_a (1 - a) = w_0 + w_a z (1 + z)^{-1}
\label{eq: cpleos}
\end{equation}
so that, in a flat universe filled by dust matter and dark energy, the
dimensionless Hubble parameter $E(z) = H/H_0$ reads\,:

\begin{equation}
E^2(z) = \Omega_M (1 + z)^3 + \Omega_X (1 + z)^{3(1 + w_0 + w_a)} {\rm
e}^{-\frac{3 w_a z}{1 + z}}
\label{eq: ecpl}
\end{equation}
with $\Omega_X = 1 - \Omega_M$ because of the flatness assumption. In order
to determine the cosmographic parameters for such a model, we avoid
integrating $H(z)$ to get $a(t)$ by noting that $d/dt = -(1 + z) H(z)
d/dz$. We can use such a relation to evaluate $(\dot{H}, \ddot{H},
\dddot{H}, d^4H/dt^4)$ and then solve Eqs.(\ref{eq: hdot})\,-\,(\ref{eq: h4dot}),
evaluated in $z = 0$, with respect to the parameters of interest. Some
algebra finally gives\,:

\begin{equation}
q_0 = \frac{1}{2} + \frac{3}{2} (1 - \Omega_M) w_0 \ ,
\label{eq: qzcpl}
\end{equation}

\begin{equation}
j_0 = 1 + \frac{3}{2} (1 - \Omega_M) \left [ 3w_0 (1 + w_0) + w_a \right ]
\ , \label{eq: jzcpl}
\end{equation}

\begin{eqnarray}
s_0 & = & -\frac{7}{2} - \frac{33}{4} (1 - \Omega_M) w_a \nonumber \\ ~ & -
& \frac{9}{4} (1 - \Omega_M) \left [ 9 + (7 - \Omega_M) w_a \right ] w_0
\nonumber \\ ~ & - & \frac{9}{4} (1 - \Omega_M) (16 - 3\Omega_M) w_0^2 \nonumber \\
~ & - & \frac{27}{4} (1 - \Omega_M) (3 - \Omega_M) w_0^3 \ ,
\label{eq: szcpl}
\end{eqnarray}

\begin{eqnarray}
l_0 & = & \frac{35}{2} + \frac{1 - \Omega_M}{4} \left [ 213 + (7 -
\Omega_M) w_a \right ] w_a \nonumber \\ ~ & + & \frac{1 - \Omega_M)}{4}
\left [ 489 + 9(82 - 21 \Omega_M) w_a \right ] w_0 \nonumber \\
~ & + & \frac{9}{2} (1 - \Omega_M) \left [ 67 - 21 \Omega_M + \frac{3}{2}
(23 - 11 \Omega_M) w_a \right ] w_0^2 \nonumber \\ ~ & + & \frac{27}{4} (1
- \Omega_M) (47 - 24 \Omega_M) w_0^3 \nonumber \\ ~ & + & \frac{81}{2} (1 - \Omega_M) (3 - 2\Omega_M) w_0^4 \ .
\label{eq: lzcpl}
\end{eqnarray}
Inserting Eqs.(\ref{eq: qzcpl})\,-\,(\ref{eq: lzcpl}) into Eqs.(\ref{eq:
f0z})\,-\,(\ref{eq: defr}), we get lengthy expressions (which we do not
report here) giving the present day values of $f(R)$ and its first three
derivatives as function of $(\Omega_M, w_0, w_a)$. It is worth noting that
the $f(R)$ model thus obtained is not dynamically equivalent to the
starting CPL one. Indeed, the two models have the same cosmographic
parameters only today. As such, for instance, the scale factor is the same
between the two theories only over the time period during which the fifth
order Taylor expansion is a good approximation of the actual $a(t)$. It is
also worth stressing that such a procedure does not select a unique $f(R)$
model, but rather a class of fourth order theories all sharing the same
third order Taylor expansion of $f(R)$.

\subsection{The $\Lambda$CDM case}

With these caveats in mind, it is worth considering first the $\Lambda$CDM
model which is obtained by setting $(w_0, w_a) = (-1, 0)$ in the above
expressions thus giving\,:

\begin{equation}
\left \{
\begin{array}{lll}
\displaystyle{q_0} & = & \displaystyle{\frac{1}{2} - \frac{3}{2} \Omega_{\Lambda}} \\
~ & ~ & \\
\displaystyle{j_0} & = & \displaystyle{1} \\
~ & ~ & \\
\displaystyle{s_0} & = & \displaystyle{1 - \frac{9}{2} \Omega_M} \\
~ & ~ & \\
\displaystyle{l_0} & = & \displaystyle{1 + 3 \Omega_M + \frac{27}{2} \Omega_M^2} \\
\end{array}
\right . \ .
\label{eq: cplcdm}
\end{equation}
When inserted into the expressions for the $f(R)$ quantities, these
relations give the remarkable result\,:

\begin{equation}
f(R_0) = R_0 + 2\Lambda \ \ , \ \ f''(R_0) = f'''(R_0) = 0 \ ,
\label{eq: frlcdm}
\end{equation}
so that we obviously  conclude that the only $f(R)$ theory having exactly
the same cosmographic parameters as the $\Lambda$CDM model is just $f(R)
\propto R$, i.e. GR. It is worth noticing that such a result comes out as
a consequence of the values of $(q_0, j_0)$ in the $\Lambda$CDM model.
Indeed, should we have left $(s_0, l_0)$ undetermined and only fixed $(q_0,
j_0)$ to the values in (\ref{eq: cplcdm}), we should have got the same
result in (\ref{eq: frlcdm}). Since the $\Lambda$CDM model fits well a
large set of different data, we do expect that the actual values of $(q_0,
j_0, s_0, l_0)$ do not differ too much from the $\Lambda$CDM ones.
Therefore, we plug into Eqs.(\ref{eq: f0z})\,-\,(\ref{eq: defr}) the
following expressions\,:

\begin{displaymath}
q_0 = q_0^{\Lambda} {\times} (1 + \varepsilon_q) \ \ , \ \ j_0 = j_0^{\Lambda} {\times}
(1 + \varepsilon_j) \ \ ,
\end{displaymath}

\begin{displaymath}
s_0 = s_0^{\Lambda} {\times} (1 + \varepsilon_s) \ \ , \ \ l_0 = l_0^{\Lambda} {\times}
(1 + \varepsilon_l) \ \ ,
\end{displaymath}
with $(q_0^{\Lambda}, j_0^{\Lambda}, s_0^{\Lambda}, l_0^{\Lambda})$ given
by Eqs.(\ref{eq: cplcdm}) and $(\varepsilon_q, \varepsilon_j,
\varepsilon_s, \varepsilon_l)$ quantifyin the deviations from the $\Lambda$CDM
values allowed by the data. A numerical estimate of these quantities may be
obtained, e.g., from a Markov chain analysis, but this is outside our aims.
Since we are here interested in a theoretical examination, we prefer to
consider an idealized situation where the four quantities above all share
the same value $\varepsilon << 1$. In such a case, we can easily
investigate how much the corresponding $f(R)$ deviates from the GR one
considering the two ratios $f''(R_0)/f(R_0)$ and $f'''(R_0)/f(R_0)$.
Inserting the above expressions for the cosmographic parameters into the
exact (not reported) formulae for $f(R_0)$, $f''(R_0)$ and $f'''(R_0)$,
taking their ratios and then expanding to first order in $\varepsilon$, we
finally get\,:

\begin{equation}
\eta_{20} = \frac{64 - 6 \Omega_M (9\Omega_M + 8)}
{\left [ 3 (9\Omega_M + 74) \Omega_M - 556 \right ] \Omega_M^2 + 16} \ {\times} \
\frac{\varepsilon}{27} \ ,
\label{eq: e20eps}
\end{equation}

\begin{equation}
\eta_{30} = \frac{6 \left [ (81 \Omega_M - 110) \Omega_M + 40 \right ] \Omega_M + 16}
{\left [ 3 (9\Omega_M + 74) \Omega_M - 556 \right ] \Omega_M^2 + 16} \ {\times} \
\frac{\varepsilon}{243 \Omega_M^2} \ ,
\label{eq: e30eps}
\end{equation}
having defined $\eta_{20} = f''(R_0)/f(R_0) {\times} H_0^4$ and $\eta_{30} =
f'''(R_0)/f(R_0) {\times} H_0^6$ which, being dimensionless quantities, are more
suited to estimate the order of magnitudes of the different terms.
Inserting our fiducial values for $\Omega_M$, we get\,:

\begin{displaymath}
\left \{
\begin{array}{ll}
\displaystyle{\eta_{20} \simeq 0.15 \ {\times} \ \varepsilon} & {\rm for}
\ \ \Omega_M = 0.041 \\
~ & ~ \\
\displaystyle{\eta_{20} \simeq -0.12 \ {\times} \ \varepsilon} & {\rm for}
\ \ \Omega_M = 0.250 \\
\end{array}
\right . \ ,
\end{displaymath}

\begin{displaymath}
\left \{
\begin{array}{ll}
\displaystyle{\eta_{30} \simeq 4 \ {\times} \ \varepsilon} & {\rm for}
\ \ \Omega_M = 0.041 \\
~ & ~ \\
\displaystyle{\eta_{30} \simeq -0.18 \ {\times} \ \varepsilon} & {\rm for}
\ \ \Omega_M = 0.250 \\
\end{array}
\right . \ .
\end{displaymath}
For values of $\varepsilon$ up to 0.1, the above relations show
that the second and third derivatives are at most two orders of
magnitude smaller than the zeroth order term $f(R_0)$. Actually,
the values of $\eta_{30}$ for a baryon only model (first row)
seems to argue in favor of a larger importance of the third order
term. However, we have numerically checked that the above
relations approximates very well the exact expressions up to
$\varepsilon \simeq 0.1$ with an accuracy depending on the value
of $\Omega_M$, being smaller for smaller matter density
parameters. Using the exact expressions for $\eta_{20}$ and
$\eta_{30}$, our conclusion on the negligible effect of the second
and third order derivatives are significantly strengthened.

Such a result holds under the hypotheses that the narrower are the
constraints on the validity of the $\Lambda$CDM model, the smaller are the
deviations of the cosmographic parameters from the $\Lambda$CDM ones. It is
possible to show that this indeed the case for the CPL parametrization we
are considering. On the other hand, we have also assumed that the
deviations $(\varepsilon_q, \varepsilon_j, \varepsilon_s, \varepsilon_l)$
take the same values. Although such hypothesis is somewhat ad hoc, we argue
that the main results are not affected by giving it away. Indeed, although
different from each other, we can still assume that all of them are very
small so that Taylor expanding to the first order should lead to additional
terms into Eqs.(\ref{eq: e20eps})\,-\,(\ref{eq: e30eps}) which are likely
of the same order of magnitude. We may therefore conclude that, if the
observations confirm that the values of the cosmographic parameters agree
within $\sim 10\%$ with those predicted for the $\Lambda$CDM model, we must
conclude that the deviations of $f(R)$ from the GR case, $f(R) \propto R$,
should be vanishingly small.

It is worth stressing, however, that such a conclusion only holds for those
$f(R)$ models satisfying the constraint (\ref{eq: checkcond}). It is indeed
possible to work out a model having $f(R_0) \propto R_0$, $f''(R_0) =
f'''(R_0) = 0$, but $f^{(n)}(R_0) \ne 0$ for some $n$. For such a (somewhat
ad hoc) model, Eq.(\ref{eq: checkcond}) is clearly not satisfied so that
the cosmographic parameters have to be evaluated from the solution of the
field equations. For such a model, the conclusion above does not hold so
that one cannot exclude that the resulting $(q_0, j_0, s_0, l_0)$ are
within $10\%$ of the $\Lambda$CDM ones.

\subsection{The constant EoS model}

Let us now take into account the condition $w = -1$, but still retains $w_a
= 0$ thus obtaining the so called {\it quiessence} models. In such a case,
some problems arise because both the terms $(j_0 - q_0 - 2)$ and
${\cal{R}}$ may vanish for some combinations of the two model parameters
$(\Omega_M, w_0)$. For instance, we find that $j_0 - q_0 - 2 = 0$  for $w_0
= (w_1, w_2)$ with\,:

\begin{displaymath}
w_1 = \frac{1}{1 - \Omega_M + \sqrt{(1 - \Omega_M) (4 - \Omega_M)}} \ ,
\end{displaymath}

\begin{displaymath}
w_2 = - \frac{1}{3} \left [ 1 + \frac{4 - \Omega_M}{\sqrt{(1 - \Omega_M) (4
- \Omega_M)}} \right ] \ .
\end{displaymath}
On the other hand, the equation ${\cal{R}}(\Omega_M, w_0) = 0$ may
have different real roots for $w$ depending on the adopted value
of $\Omega_M$. Denoting collectively with ${\bf w}_{null}$ the
values of $w_0$ that, for a given $\Omega_M$, make $(j_0 - q_0 -
2) {\cal{R}}(\Omega_M, w_0)$ taking the null value, we individuate
a set of quiessence models whose cosmographic parameters give rise
to divergent values of $f(R_0$, $f''(R_0)$ and $f'''(R_0)$. For
such models, $f(R)$ is clearly not defined so that we have to
exclude these cases from further consideration. We only note that
it is still possible to work out a $f(R)$ theory reproducing the
same background dynamics of such models, but a different route has
to be used.

\begin{figure}
\centering \resizebox{8.5cm}{!}{\includegraphics{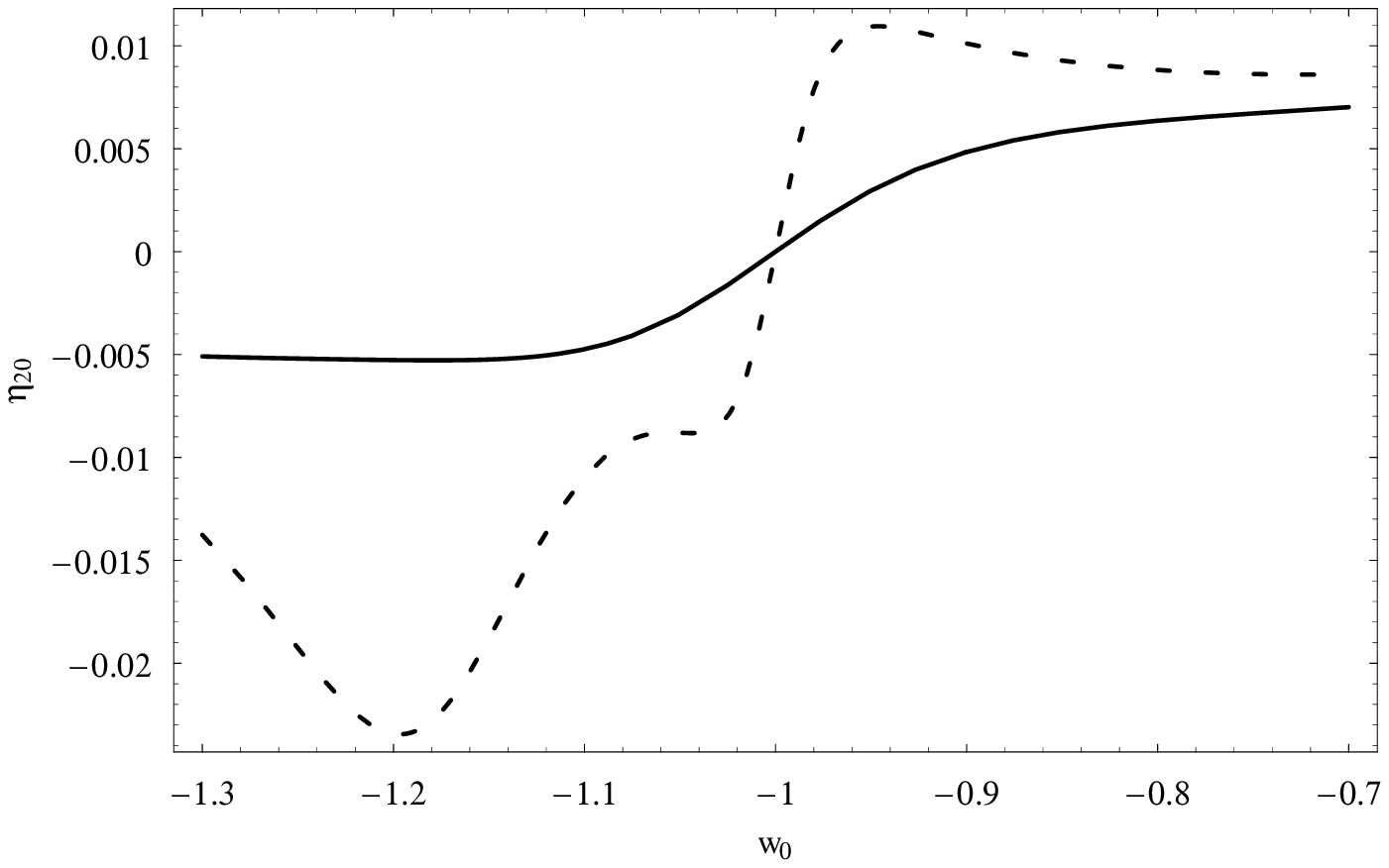}}
\caption{The dimensionless ratio between the present day values of $f''(R)$ and $f(R)$
as function of the constant EoS $w_0$ of the corresponding quiessence
model. Short dashed and solid lines refer to models with $\Omega_M = 0.041$
and $0.250$ respectively.}
\label{fig: r20}
\end{figure}

Since both $q_0$ and $j_0$ now deviate from the $\Lambda$CDM values, it is
not surprising that both $f''(R_0)$ and $f'''(R_0)$ take finite non null
values. However, it is more interesting to study the two quantities
$\eta_{20}$ and $\eta_{30}$ defined above to investigate the deviations of
$f(R)$ from the GR case. These are plotted in Figs.\,\ref{fig: r20} and
\ref{fig: r30} for the two fiducial $\Omega_M$ values. Note that the range
of $w_0$ in these plots have been chosen in order to avoid divergences, but
the lessons we will draw also hold for the other $w_0$ values.

As a general comment, it is clear that, even in this case,
$f''(R_0)$ and $f'''(R_0)$ are from two to three orders of
magnitude smaller that the zeroth order term $f(R_0)$. Such a
result could be yet guessed from the previous discussion for the
$\Lambda$CDM case. Actually, relaxing the hypothesis $w_0 = -1$ is
the same as allowing the cosmographic parameters to deviate from
the $\Lambda$CDM values. Although a direct mapping between the two
cases cannot be established, it is nonetheless evident that such a
relation can be argued thus making the outcome of the above plots
not fully surprising. It is nevertheless worth noting that, while
in the $\Lambda$CDM case, $\eta_{20}$ and $\eta_{30}$ always have
opposite signs, this is not the case for quiessence models with $w
> -1$. Indeed, depending on the value of $\Omega_M$, we can have
$f(R)$ theories with both $\eta_{20}$ and $\eta_{30}$ positive.
Moreover, the lower is $\Omega_M$, the higher are the ratios
$\eta_{20}$ and $\eta_{30}$ for a given value of $w_0$. This can
be explained qualitatively noticing that, for a lower $\Omega_M$,
the density parameter of the curvature fluid (playing the role of
an effective dark energy) must be larger thus claiming for higher
values of the second and third derivatives (see also \cite{emilio}
for a different approach to the problem).

\subsection{The general case}

Finally, we consider evolving dark energy models with $w_a \ne 0$.
Needless to say, varying three parameters allows to get a wide
range of models that cannot be discussed in detail. Therefore, we
only concentrate on evolving dark energy models with $w_0 = -1$ in
agreement with some most recent analysis. The results on
$\eta_{20}$ and $\eta_{30}$ are plotted in Figs.\,\ref{fig:
r20cpl} and \ref{fig: r30cpl} where these quantities as functions
of $w_a$. Note that we are considering models with positive $w_a$
so that $w(z)$ tends to $w_0 + w_a > w_0$ for $z \rightarrow
\infty$ so that the EoS dark energy can eventually approach the
dust value $w = 0$. Actually, this is also the range favored by the data.
We have, however, excluded values where $\eta_{20}$ or $\eta_{30}$ diverge.
Considering how they are defined, it is clear that these two quantities
diverge when $f(R_0) = 0$ so that the values of $(w_0, w_a)$ making
$(\eta_{20}, \eta_{30})$ to diverge may be found solving\,:

\begin{figure}
\centering
\resizebox{8.5cm}{!}{\includegraphics{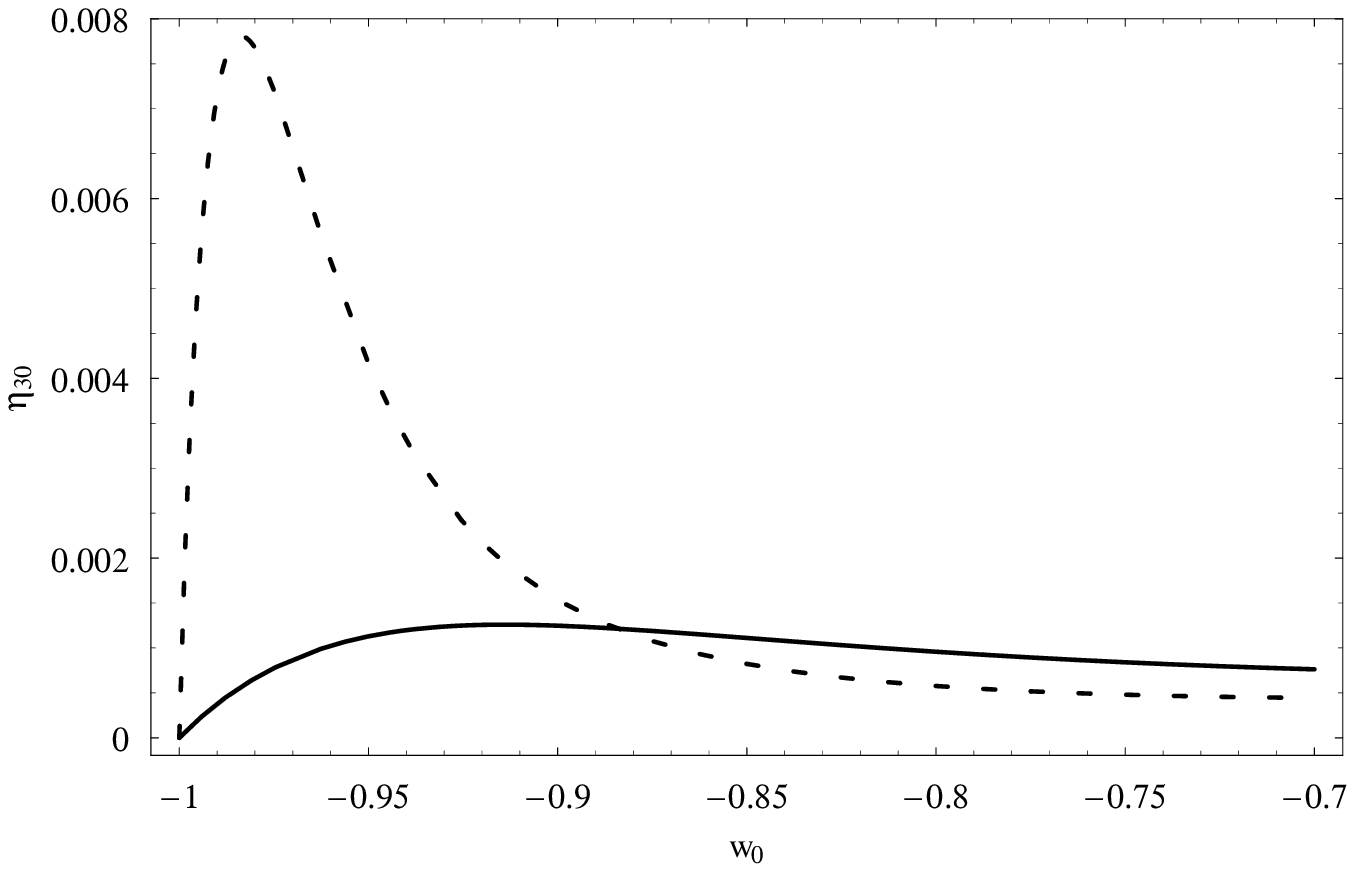}}
\caption{The dimensionless ratio between the present day values of
$f'''(R)$ and $f(R)$ as function of the constant EoS $w_0$ of the
corresponding quiessence model. Short dashed and solid lines refer
to models with $\Omega_M = 0.041$ and $0.250$ respectively.}
\label{fig: r30}
\end{figure}

\begin{displaymath}
{\cal{P}}_0(w_0, w_a) \Omega_M + {\cal{Q}}_0(w_0, w_a) = 0
\end{displaymath}
where ${\cal{P}}_0(w_0, w_a)$ and ${\cal{Q}}_0(w_0, w_a)$ are obtained by
inserting Eqs.(\ref{eq: qzcpl})\,-\,(\ref{eq: lzcpl}) into the defintions
(\ref{eq: defp0})\,-\,(\ref{eq: defq0}). For such CPL models, there is no
any $f(R)$ model having the same cosmographic parameters and, at the same
time, satisfying all the criteria needed for the validity of our procedure.
Actually, if $f(R_0) = 0$, the condition (\ref{eq: checkcond}) is likely to
be violated so that higher than third order must be included in the Taylor
expansion of $f(R)$ thus invalidating the derivation of Eqs.(\ref{eq:
f0z})\,-\,(\ref{eq: f3z}).

Under these caveats, Figs.\,\ref{fig: r20cpl} and \ref{fig: r30cpl}
demonstrate that allowing the dark energy EoS to evolve does not change
significantly our conclusions. Indeed, the second and third derivatives,
although being not null, are nevertheless negligible with respect to the
zeroth order term thus arguing in favour of a GR\,-\,like $f(R)$ with only
very small corrections. Such a result is, however, not fully unexpected.
From Eqs.(\ref{eq: qzcpl}) and (\ref{eq: jzcpl}), we see that, having
setted $w_0 = -1$, the $q_0$ parameter is the same as for the $\Lambda$CDM
model, while $j_0$ reads $j_0^{\Lambda} + (3/2)(1 - \Omega_M) w_a$. As we
have stressed above, the Hilbert - Einstein Lagrangian $f(R) = R + 2
\Lambda$ is recovered when $(q_0, j_0) = (q_0^{\Lambda},
j_0^{\Lambda})$ whatever the values of $(s_0, l_0)$ are. Introducing a $w_a
\ne 0$ makes $(s_0, l_0)$ to differ from the $\Lambda$CDM values, but the
first two cosmographic parameters are only mildly affected. Such deviations
are then partially washed out by the complicated way they enter in the
determination of the present day values of $f(R)$ and its first three
derivatives.

\begin{figure}
\centering \resizebox{8.5cm}{!}{\includegraphics{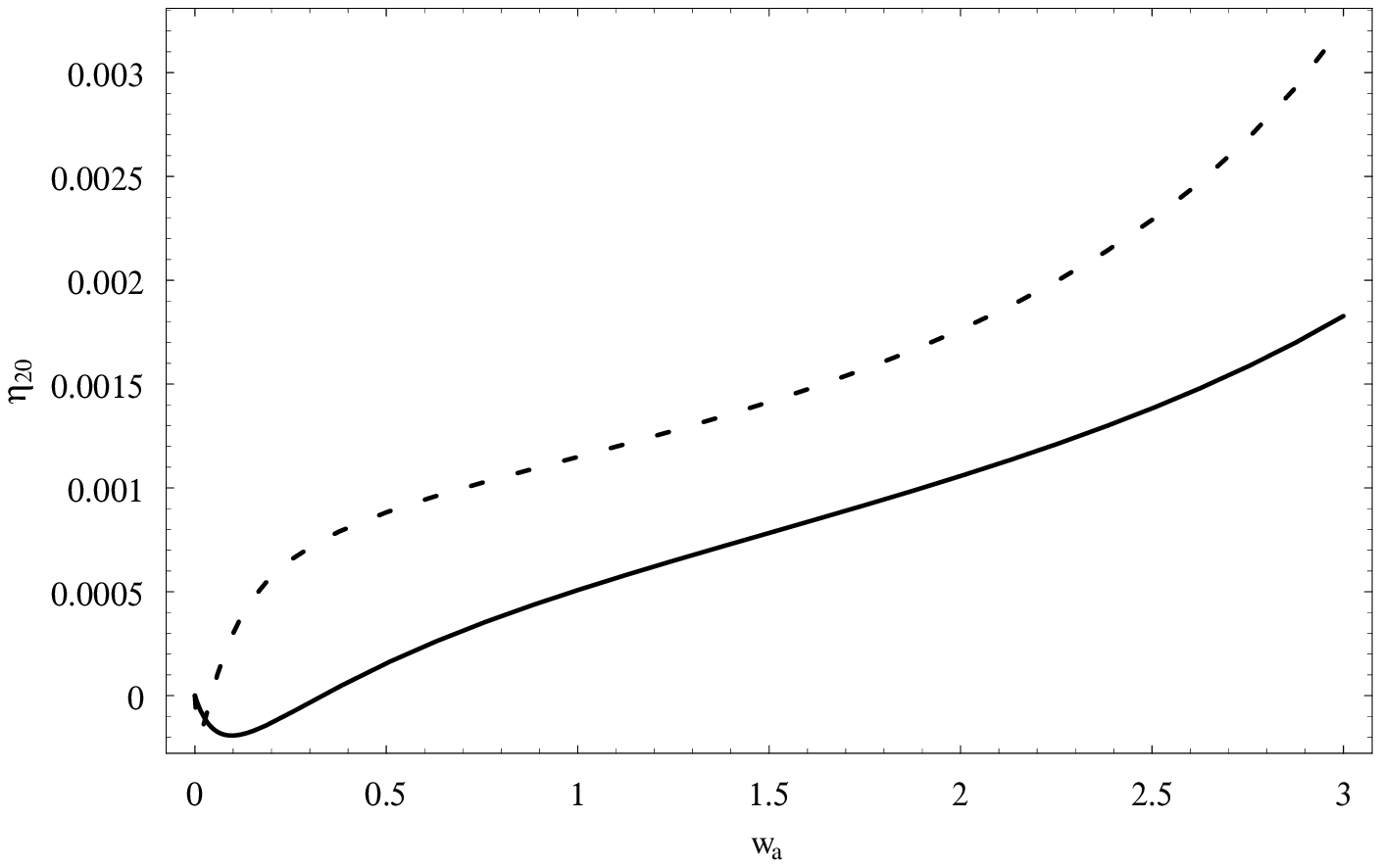}}
\caption{The dimensionless ratio between the present day values of $f''(R)$ and $f(R)$
as function of the $w_a$ parameter for models with $w_0 = -1$. Short dashed
and solid lines refer to models with $\Omega_M = 0.041$ and $0.250$
respectively.}
\label{fig: r20cpl}
\end{figure}

\section{Constraining $f(R)$ parameters}

In the previous section, we have worked an alternative method to
estimate $f(R_0)$, $f''(R_0)$, $f'''(R_0)$ resorting to a model
independent parameterization of the dark energy EoS. However, in
the ideal case, the cosmographic parameters are directly estimated
from the data so that Eqs.(\ref{eq: f0z})\,-\,(\ref{eq: defr}) can
be used to infer the values of the $f(R)$ related quantities.
These latter can then be used to put constraints on the parameters
entering an assumed fourth order theory assigned by a $f(R)$
function characterized by a set of parameters ${\bf p} = (p_1,
\ldots, p_n)$ provided that the hypotheses underlying the
derivation of Eqs.(\ref{eq: f0z})\,-\,(\ref{eq: defr}) are indeed
satisfied. We show below two interesting cases which clearly
highlight the potentiality and the limitations of such an
analysis.

\subsection{Double power law Lagrangian}

As a first interesting example, we set\,:

\begin{equation}
f(R) = R \left (1 + \alpha R^{n} + \beta R^{-m} \right )
\label{eq: frdpl}
\end{equation}
with $n$ and $m$ two positive real numbers (see, for example,
\cite{double} for some physical motivations). The following
expressions are immediately obtained\,:

\begin{displaymath}
\left \{
\begin{array}{lll}
f(R_0) & = & R_0 \left (1 + \alpha R_0^{n} + \beta R_0^{-m} \right
) \\ ~ & ~ & ~ \\ f'(R_0) & = & 1 + \alpha (n + 1) R_0^n - \beta
(m - 1) R_0^{-m} \\  ~ & ~ & ~ \\ f''(R_0) & = & \alpha n (n + 1)
R_0^{n - 1} + \beta m (m - 1) R_0^{-(1 + m)} \\ ~ & ~ & ~ \\
f'''(R_0) & = & \alpha n (n + 1) (n - 1) R_0^{n - 2} \\ ~ & - &
\beta m (m + 1) (m - 1) R_0^{-(2 + m)}
\end{array}
\right . \ .
\label{eq: derdpl}
\end{displaymath}
Denoting by $\phi_i$ (with $i = 0, \ldots, 3$) the values of
$f^{(i)}(R_0)$ determined through Eqs.(\ref{eq:
f0z})\,-\,(\ref{eq: defr}), we can solve\,:

\begin{figure}
\centering \resizebox{8.5cm}{!}{\includegraphics{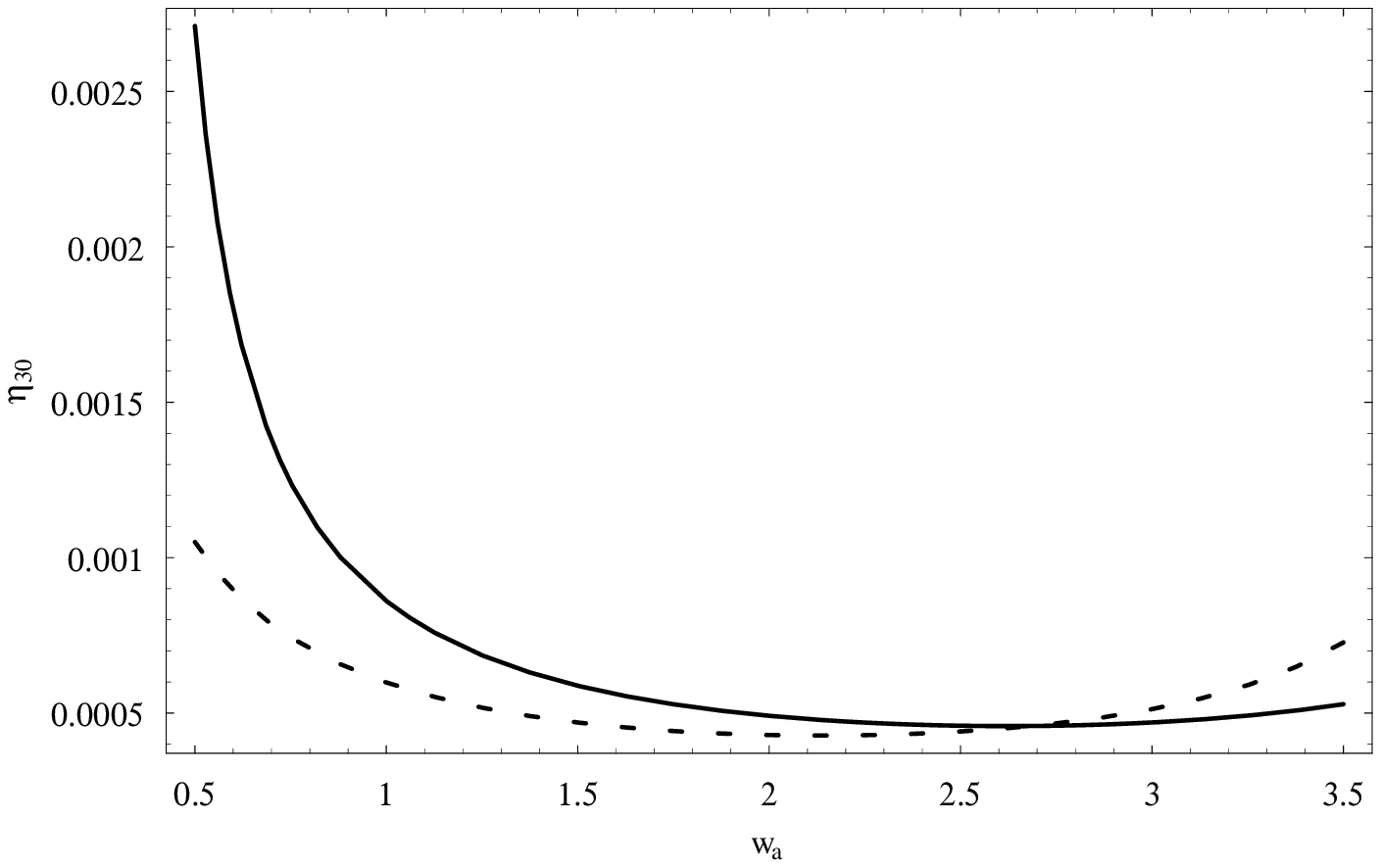}}
\caption{The dimensionless ratio between the present day values of $f'''(R)$ and $f(R)$
as function of the $w_a$ parameter for models with $w_0 = -1$. Short dashed
and solid lines refer to models with $\Omega_M = 0.041$ and $0.250$
respectively. }
\label{fig: r30cpl}
\end{figure}

\begin{displaymath}
\left \{
\begin{array}{lll}
f(R_0) & = & \phi_0 \\ f'(R_0) & = & \phi_1 \\ f''(R_0) & = & \phi_2 \\
f'''(R_0) & = & \phi_3 \\
\end{array}
\right .
\end{displaymath}
which is a system of four equations in the four unknowns $(\alpha,
\beta, n, m)$ that can be analytically solved proceeding as
follows. First, we solve the first and second equation with
respect to $(\alpha, \beta)$ obtaining\,:

\begin{equation}
\left \{
\begin{array}{lll}
\alpha & = & \displaystyle{\frac{1 - m}{n + m} \left ( 1 - \frac{\phi_0}{R_0} \right ) R_0^{-n}} \\
~ & ~ & ~ \\ \beta & = & \displaystyle{- \frac{1 + n}{n + m} \left ( 1 -
\frac{\phi_0}{R_0} \right ) R_0^{m}} \\
\end{array}
\right . \ ,
\label{eq: ab12}
\end{equation}
while, solving the third and fourth equations, we get\,:

\begin{equation}
\left \{
\begin{array}{lll}
\alpha & = & \displaystyle{\frac{\phi_2 R_0^{1 - n} \left [ 1 + m + (\phi_3/\phi_2) R_0 \right ]}{n (n + 1) (n + m)}} \\
~ & ~ & ~ \\
\beta & = & \displaystyle{\frac{\phi_2 R_0^{1 + n} \left [ 1 - n + (\phi_3/\phi_2) R_0 \right ]}{m (1 - m) (n + m)}} \\
\end{array}
\right . \ .
\label{eq: ab34}
\end{equation}
Equating the two solutions, we get a systems of two equations in the two
unknowns $(n, m)$, namely\,:

\begin{equation}
\left \{
\begin{array}{lll}
\displaystyle{\frac{n (n + 1) (1 - m) \left ( 1 - \phi_0/R_0 \right )}
{\phi_2 R_0 \left [ 1 + m + (\phi_3/\phi_2) R_0 \right ]}} & = & 1 \\ ~ & ~
& ~ \\ \displaystyle{\frac{m (n + 1) (m - 1) \left ( 1 - \phi_0/R_0 \right
)} {\phi_2 R_0 \left [ 1 - n + (\phi_3/\phi_2) R_0 \right ]}} & = & 1
\end{array}
\right . \ .
\end{equation}
Solving with respect to $m$, we get two solutions, the first one being $m =
-n$ which has to be discarded since makes $(\alpha, \beta)$ goes to infinity. The only
acceptable solution is\,:

\begin{equation}
m = - \left [ 1 - n + (\phi_3/\phi_2) R_0 \right ]
\label{eq: msol}
\end{equation}
which, inserted back into the above system, leads to a second order
polynomial equation for $n$ with solutions\,:

\begin{equation}
n = \frac{1}{2} \left [1 + \frac{\phi_3}{\phi_2} R_0 {\pm}
\frac{\sqrt{{\cal{N}}(\phi_0, \phi_2, \phi_3)}}{\phi_2 R_0 (1 + \phi_0/R_0)} \right ]
\label{eq: nsol}
\end{equation}
where we have defined\,:

\begin{eqnarray}
{\cal{N}}(\phi_0, \phi_2, \phi_3) & = & \left ( R_0^2 \phi_0^2 - 2 R_0^3
\phi_0 + R_0^4 \right ) \phi_3^2 \nonumber \\
~ & + & 6 \left ( R_0 \phi_0^2 - 2 R_0^2 \phi_0 + R_0^3 \right ) \phi_2
\phi_3 \nonumber \\
~ & + & 9 \left ( \phi_0^2 - 2 R_0 \phi_0 + R_0^2 \right ) \phi_2^2
\nonumber \\ ~ & + & 4 \left ( R_0^2 \phi_0 - R_0^3 \right ) \phi_2^3 \ .
\label{eq: defn}
\end{eqnarray}
Depending on the values of $(q_0, j_0, s_0, l_0)$, Eq.(\ref{eq:
nsol}) may lead to one, two or any acceptable solution, i.e. real
positive values of $n$. This solution has then to be inserted back
into Eq.(\ref{eq: msol}) to determine $m$ and then into
Eqs.(\ref{eq: ab12}) or (\ref{eq: ab34}) to estimate $(\alpha,
\beta)$. If the final values of $(\alpha, \beta, n, m)$  are
physically viable, we can conclude that the model in Eq.(\ref{eq:
frdpl}) is in agreement with the data giving the same cosmographic
parameters inferred from the data themselves. Exploring
analytically what is the region of the $(q_0, j_0, s_0, l_0)$
parameter space which leads to acceptable $(\alpha, \beta, n, m)$
solutions is a daunting task far outside the aim of the present
paper.

\subsection{HS model}

One of the most pressing problems of $f(R)$ theories is the need to escape
the severe constraints imposed by the Solar System tests. A successful
model has been recently proposed by Hu and Sawicki \cite{Hu} (HS)
setting\footnote{Note that such a model does not pass the matter
instability test so that some viable generalizations \cite{Odi} have been
proposed.}\,:

\begin{equation}
f(R) = R - R_c \frac{\alpha (R/R_c)^n}{1 + \beta (R/R_c)^n} \ .
\label{eq: frhs}
\end{equation}
As for the double power law model discussed above, there are four
parameters which we can be expressed in terms of the cosmographic
parameters $(q_0, j_0, s_0, l_0)$.

As a first step, it is trivial to get\,:

\begin{equation}
\left \{
\begin{array}{lll}
f(R_0) & = & \displaystyle{R_0 - R_c \frac{\alpha R_{0c}^n}{1 +
\beta R_{0c}^n}} \\  ~ & ~ & ~ \\  f'(R_0) & = & \displaystyle{1
- \frac{\alpha n R_c R_{0c}^{n}}{R_0 (1 + \beta R_{0c}^n)^2}} \\
~ & ~ & ~ \\ f''(R_0) & = & \displaystyle{\frac{\alpha n R_c
R_{0c}^n \left [ (1 - n) + \beta (1 + n) R_{0c}^n \right ]}{R_0^2
(1 + \beta R_{0c}^n)^3}} \\ ~ & ~ & ~ \\ f'''(R_0) & = &
\displaystyle{\frac{\alpha n R_c R_{0c}^n (A n^2 + B n + C)}{R_0^3
(1 + \beta R_{0c}^n)^4}}
\end{array}
\right .
\label{eq: derhs}
\end{equation}
with $R_{0c} = R_0/R_c$ and\,:

\begin{equation}
\left \{
\begin{array}{lll}
A & = & - \beta^2 R_{0c}^{2n} + 4 \beta R_{0c}^n - 1 \\  ~ & ~ & ~ \\
B & = & 3 (1 - \beta^2  R_{0c}^{2n}) \\ ~ & ~ & ~ \\ C & = &  -2
(1 - \beta  R_{0c}^{n})^2
\end{array}
\right . \ .
\end{equation}
Equating Eqs.(\ref{eq: derhs}) to the four quantities $(\phi_0,
\phi_1, \phi_2, \phi_3)$ defined as above, we could, in principle,
solve this system of four equations in four unknowns to get
$(\alpha, \beta, R_c, n)$ in terms of $(\phi_0, \phi_1, \phi_2,
\phi_3)$ and then, using Eqs.(\ref{eq: f0z})\,-\,(\ref{eq: defr})
as functions of the cosmographic parameters. However, setting $\phi_1 = 1$
as required by Eq.(\ref{eq: f1z}) gives the only trivial solution $\alpha n
R_c = 0$ so that the HS model reduces to the Einstein\,-\,Hilbert
Lagrangian $f(R) = R$. In order to escape this problem, we can relax the
condition $f'(R_0) = 1$ to $f'(R_0) = (1 + \varepsilon)^{-1}$.  As we have
discussed in Sect.\,IV, this is the same as assuming that the present day
effective gravitational constant $G_{eff, 0} = G_N/f'(R_0)$ only slightly
differs from the usual Newtonian one which seems to be a quite reasonable
assumption. Under this hypothesis, we can analytically solve for $(\alpha,
\beta, R_c, n)$ in terms of $(\phi_0, \varepsilon, \phi_2, \phi_3)$. The
actual values of $(\phi_0, \phi_2, \phi_3)$ will be no more given by
Eqs.(\ref{eq: f0z})\,-\,(\ref{eq: f3z}), but we have checked that they
deviate from those expressions\footnote{Note that the correct expressions
for $(phi_0, \phi_2, \phi_3)$ may still formally be written as
Eqs.(\ref{eq: f0z})\,-\,(\ref{eq: f3z}), but the polynomials entering them
are now different and also depend on powers of $\varepsilon$.} much less
than $10\%$ for $\varepsilon$ up to $10\%$ well below any realistic
expectation.

With this caveat in mind, we first solve

\begin{displaymath}
f(R_0) = \phi_0 \ \ , \ \ f''(R_0) = (1 + \varepsilon)^{-1}
\end{displaymath}
to get\,:

\begin{eqnarray}
\alpha & = & \frac{n (1 + \varepsilon)}{\varepsilon} \left (
\frac{R_0}{R_c} \right )^{1 - n} \left ( 1 - \frac{\phi_0}{R_0} \right )^2 \ , \nonumber \\
\beta & = & \frac{n (1 + \varepsilon)}{\varepsilon} \left (
\frac{R_0}{R_c} \right )^{-n} \left [ 1  - \frac{\phi_0}{R_0}  -
\frac{\varepsilon}{n (1 + \varepsilon)}\right ] \ . \nonumber
\end{eqnarray}
Inserting these expressions in Eqs.(\ref{eq: derhs}), it is easy to check
that $R_c$ cancels out so that we can no more determine its value. Such a
result is, however, not unexpected. Indeed, Eq.(\ref{eq: frhs}) can
trivially be rewritten as\,:

\begin{displaymath}
f(R) = R - \frac{\tilde{\alpha} R^n}{1 + \tilde{\beta} R^n}
\end{displaymath}
with $\tilde{\alpha} = \alpha R_c^{1 - n}$ and $\tilde{\beta} =
\beta R_c^{-n}$ which are indeed the quantities that are
determined by the above expressions for $(\alpha, \beta)$.
Reversing the discussion, the present day values of $f^{(i)}(R)$
depend on $(\alpha, \beta, R_c)$ only through the two parameters
$(\tilde{\alpha}, \tilde{\beta})$. As such, the use of
cosmographic parameters is unable to break this degeneracy.
However, since $R_c$ only plays the role of a scaling parameter,
we can arbitrarily set its value without loss of generality.

On the other hand, this degeneracy allows us to get a consistency relation
to immediately check whether the HS model is viable or not. Indeed, solving
the equation $f''(R_0) = \phi_2$, we get\,:

\begin{displaymath}
n = \frac{(\phi_0/R_0) + [(1 + \varepsilon)/\varepsilon](1 -
\phi_2 R_0) - (1 - \varepsilon)/(1 + \varepsilon)}{1 - \phi_0/R_0}
\ ,
\end{displaymath}
which can then be inserted into the equations $f'''(R_0) = \phi_3$
to obtain a complicated relation among $(\phi_0, \phi_2, \phi_3)$
which we do not report for sake of shortness. Solving such a
relation with respect to $\phi_3/\phi_0$ and Taylor expanding to
first order in $\varepsilon$, the constraint we get reads\,:

\begin{displaymath}
\frac{\phi_3}{\phi_0} \simeq - \frac{1 + \varepsilon}{\varepsilon}
\frac{\phi_2}{R_0} \left [ R_0 \left ( \frac{\phi_2}{\phi_0}
\right ) + \frac{\varepsilon \phi_0^{-1}}{1 + \varepsilon} \left (
1 - \frac{2 \varepsilon}{1 - \phi_0/R_0} \right ) \right ] \ .
\end{displaymath}
If the cosmographic parameters $(q_0, j_0, s_0, l_0)$ are known
with sufficient accuracy, one could compute the values of $(R_0,
\phi_0, \phi_2. \phi_3)$ for a given $\varepsilon$ (eventually
using the expressions obtained for $\varepsilon = 0$) and then
check if they satisfied this relation. If this is not  the case,
one can immediately give off the HS model also without the need of
solving the field equations and fitting the data. Actually, given
the still large errors on the cosmographic parameters, such a test
only remains in the realm of (quite distant) future applications.
However, the HS model works for other tests as shown in \cite{Hu}
and so a consistent cosmography analysis has to be combined with
them.

\section{Constraints on $f(R)$ derivatives from the data}

Eqs.(\ref{eq: f0z})\,-\,(\ref{eq: defr}) relate the present day values of
$f(R)$ and its first three derivatives to the cosmographic parameters
$(q_0, j_0, s_0, l_0)$ and the matter density $\Omega_M$. In principle,
therefore, a measurement of these latter quantities makes it possible to
put constraints on $f^{(i)}(R_0)$, with $i = \{0, \ldots, 3\}$, and hence
on the parameters of a given fourth order theory through the method shown
in the previous section. Actually, the cosmographic parameters are affected
by errors which obviously propagate onto the $f(R)$ quantities. Actually,
the covariance matrix for the cosmographic parameters is not diagonal so
that one has also take care of this to estimate the final errors on
$f^{(i)}(R_0)$. A similar discussion also holds for the errors on the
dimensionless ratios $\eta_{20}$ and $\eta_{30}$ introduced above. As a
general rule, indicating with $g(\Omega_M, {\bf p})$ a generic $f(R)$
related quantity depending on $\Omega_M$ and the set of cosmographic
parameters ${\bf p}$, its uncertainty reads\,:

\begin{equation}
\sigma_{g}^2 = \left | \frac{\partial g}{\partial
\Omega_M} \right |^2 \sigma_{M}^2 + \sum_{i = 1}^{i = 4}{
\left | \frac{\partial g}{\partial p_i} \right |^2
\sigma_{p_i}^2} + \sum_{i \neq j}{2 \frac{\partial g}{\partial p_i}
\frac{\partial g}{\partial p_j} C_{ij}}
\label{eq: error}
\end{equation}
where $C_{ij}$ are the elements of the covariance matrix (being $C_{ii} =
\sigma_{p_i}^2$), we have set $(p_1, p_2, p_3, p_4) = (q_0, j_0, s_0, l_0)$.
and assumed that the erorr $\sigma_M$ on $\Omega_M$ is uncorrelated with
those on ${\bf p}$. Note that this latter assumption strictly holds if the
matter density parameter is estimated from an astrophysical method (such as
estimating the total matter in the universe from the estimated halo mass
function). Alternatively, we will assume that $\Omega_M$ is constrained by
the CMBR related experiments. Since these latter mainly probes the very
high redshift universe ($z \simeq z_{lss} \simeq 1089$), while the
cosmographic parameters are concerned with the present day cosmo, one can
argue that the determination of $\Omega_M$ is not affected by the details
of the model adopted for describing the late universe. Indeed, we can
reasonably assume that, whatever is the dark energy candidate or $f(R)$
theory, the CMBR era is well approximated by the standard GR with a model
comprising only dust matter. As such, we will make the simplifying (but
well motivated) assumption that $\sigma_M$ may be reduced to very small
values and is uncorrelated with the cosmographic parameters.

Under this assumption, the problem of estimating the errors on $g(\Omega_M,
{\bf p})$ reduces to estimating the covariance matrix for the cosmographic
parameters given the details of the data set used as observational
constraints. We address this issue by computing the Fisher information
matrix (see, e.g., \cite{Teg97} and references therein) defined as\,:

\begin{equation}
F_{ij} = \left \langle \frac{\partial^2 L}{\partial \theta_i \partial
\theta_j} \right \rangle
\label{eq: deffij}
\end{equation}
with $L = -2 \ln{{\cal{L}}(\theta_1, \ldots, \theta_n)}$,
${\cal{L}}(\theta_1, \ldots, \theta_n)$ the likelihood of the experiment,
$(\theta_1, \ldots, \theta_n)$ the set of parameters to be constrained, and
$\langle \ldots \rangle$ denotes the expectation value. Actually, the
expectation value is computed by evaluating the Fisher matrix elements for
fiducial values of the model parameters $(\theta_1, \ldots, \theta_n)$,
while the covariance matrix ${\bf C}$ is finally obtained as the inverse of
${\bf F}$.

A key ingredient in the computation of ${\bf F}$ is the definition of the
likelihood which depends, of course, of what experimental constraint one is
using. To this aim, it is worth remembering that our analysis is based on
fifth order Taylor expansion of the scale factor $a(t)$ so that we can only
rely on observational tests probing quantities that are well described by
this truncated series. Moreover, since we do not assume any particular
model, we can only characterize the background evolution of the universe,
but not its dynamics which, being related to the evolution of
perturbations, unavoidably need the specification of a physical model. As a
result, the SNeIa Hubble diagram is the ideal test\footnote{See the
conclusions for further discussion on this issue.} to constrain the
cosmographic parameters. We therefore defined the likelihood as\,:

\begin{equation}
\begin{array}{l}
{\cal{L}}(H_0, {\bf p}) \propto \exp{-\chi^2(H_0, {\bf p})/2} \\ ~ \\
\chi^2(H_0, {\bf p}) = \sum_{n = 1}^{{\cal{N}}_{SNeIa}}{\displaystyle{\left [
\frac{\mu_{obs}(z_i) - \mu_{th}(z_n, H_0, {\bf p})}{\sigma_i(z_i)} \right ]^2}}
\end{array} \ ,
\label{eq: deflike}
\end{equation}
where the distance modulus to redshift $z$ reads\,:

\begin{equation}
\mu_{th}(z, H_0, {\bf p}) = 25 + 5 \log{(c/H_0)} + 5 \log{d_L(z, {\bf p})} \ ,
\label{eq: defmuth}
\end{equation}
and $d_L(z)$ is the Hubble free luminosity distance\,:

\begin{equation}
d_L(z) = (1 + z) \int_{0}^{z}{\frac{dz}{H(z)/H_0}} \ .
\label{eq: defdlhf}
\end{equation}
Using the fifth order Taylor expansion of the scale factor, we get for
$d_L(z, {\bf p})$ an analytical expression (reported in Appendix A) so that
the computation of $F_{ij}$ does not need any numerical integration (which
makes the estimate faster). As a last ingredient, we need to specify the
details of the SNeIa survey giving the redshift distribution of the sample
and the error on each measurement. Following \cite{Kim}, we
adopt\footnote{Note that, in \cite{Kim}, the authors assume the data are
separated in redshift bins so that the error becomes $\sigma^2 =
\sigma_{sys}^2/{\cal{N}}_{bin} + {\cal{N}}_{bin} (z/z_{max})^2 \sigma_m^2$
with ${\cal{N}}_{bin}$ the number of SNeIa in a bin. However, we prefer to
not bin the data so that ${\cal{N}}_{bin} = 1$.}\,:

\begin{displaymath}
\sigma_(z) = \sqrt{\sigma_{sys}^2 + \left ( \frac{z}{z_{max}} \right )^2 \sigma_m^2}
\end{displaymath}
with $z_{max}$ the maximum redshift of the survey, $\sigma_{sys}$ an
irreducible scatter in the SNeIa distance modulus and $\sigma_m$ to be
assigned depending on the photometric accuracy.

In order to run the Fisher matrix calculation, we have to set a fiducial
model which we set according to the $\Lambda$CDM predictions for the
cosmographic parameters. For $\Omega_M = 0.3$ and $h = 0.72$ (with $h$ the
Hubble constant in units of $100 {\rm km/s/Mpc}$), we get\,:

\begin{displaymath}
(q_0, j_0, s_0, l_0) = (-0.55, 1.0, -0.35, 3.11) \ .
\end{displaymath}
As a first consistency check, we compute the Fisher matrix for a survey
mimicking the recent database in \cite{D07} thus setting
$({\cal{N}}_{SNeIa}, \sigma_m) = (192, 0.33)$. After marginalizing over $h$
(which, as well known, is fully degenerate with the SNeIa absolute
magnitude ${\cal{M}}$), we get for the uncertainties\,:

\begin{displaymath}
(\sigma_1, \sigma_2, \sigma_3, \sigma_4) = (0.38, 5.4, 28.1, 74.0)
\end{displaymath}
where we are still using the indexing introduced above for the cosmographic
parameters. These values compare reasonably well with those obtained from a
cosmographic fitting of the Gold SNeIa dataset\footnote{Actually, such
estimates have been obtained computing the mean and the standard deviation
from the marginalized likelihoods of the cosmographic parameters. As such,
the central values do not represent exactly the best fit model, while the
standard deviations do not give a rigorous description of the error because
the marginalized likelihoods are manifestly non - Gaussian. Nevertheless,
we are mainly interested in an order of magnitude estimate so that we do
not care about such statistical details.} \cite{John}\,:

\begin{displaymath}
q_0 = -0.90 {\pm} 0.65 \ \ , \ \ j_0 = 2.7 {\pm} 6.7 \ \ ,
\end{displaymath}

\begin{displaymath}
s_0 = 36.5 {\pm} 52.9 \ \ , \ \ l_0 = 142.7 {\pm} 320 \ \ .
\end{displaymath}
Because of the Cramer\,-\,Rao theorem, the Fisher matrix approach
is known to provide the minimum variance errors a given experiment
can attain thus giving higher limits to its accuracy on the
determination of a set of parameters. This is indeed the case with
the comparison suggesting that our predictions are quite
optimistic. It is worth stressing, however, that the analysis in
\cite{John} used the Gold dataset which is poorer in high $z$
SNeIa than the \cite{D07} one we are mimicking so that larger
errors on the higher order parameters $(s_0, l_0)$ are expected.

Rather than computing the errors on $f(R_0)$ and its first three
derivatives, it is more interesting to look at the precision attainable on
the dimensionless ratios $(\eta_{20}, \eta_{30}$ introduced above since
they quantify how much deviations from the linear order are present. For
the fiducial model we are considering, both $\eta_{20}$ and $\eta_{30}$
vanish, while, using the covariance matrix for a present day survey and
setting $\sigma_M/\Omega_M \simeq 10\%$, their uncertainties read\,:

\begin{displaymath}
(\sigma_{20}, \sigma_{30}) = (0.04, 0.04) \ .
\end{displaymath}
As an application, we can look at Figs.\,\ref{fig: r20} and \ref{fig: r30}
showing how $(\eta_{20}, \eta_{30})$ depend on the present day EoS $w_0$
for $f(R)$ models sharing the same cosmographic parameters of a dark energy
model with constant EoS. As it is clear, also considering only the $1
\sigma$ range, the full region plotted is allowed by such large constraints on
$(\eta_{20}, \eta_{30})$ thus meaning that the full class of corresponding
$f(R)$ theories is viable. As a consequence, we may conclude that the
present day SNeIa data are unable to discriminate between a $\Lambda$
dominated universe and this class of fourth order gravity theories.

As a next step, we consider a SNAP\,-\,like survey \cite{SNAP} thus setting
$({\cal{N}}_{SNeIa}, \sigma_m) = (2000, 0.02)$. We use the same redshift
distribution in Table 1 of \cite{Kim} and add 300 nearby SNeIa in the
redshift range $(0.03, 0.08)$. The Fisher matrix calculation gives for the
uncertainties on the cosmographic parameters\,:

\begin{displaymath}
(\sigma_1, \sigma_2, \sigma_3, \sigma_4) = (0.08, 1.0, 4.8, 13.7) \ .
\end{displaymath}
The significant improvement of the accuracy in the determination of $(q_0,
j_0, s_0, l_0)$ translates in a reduction of the errors on $(\eta_{20},
\eta_{30})$ which now read\,:

\begin{displaymath}
(\sigma_{20}, \sigma_{30}) = (0.007, 0.008)
\end{displaymath}
having assumed that, when SNAP data will be available, the matter density
parameter $\Omega_M$ has been determined with a precision
$\sigma_M/\Omega_M \sim 1\%$. Looking again at Figs.\,\ref{fig: r20} and
\ref{fig: r30}, it is clear that the situation is improved. Indeed, the
constraints on $\eta_{20}$ makes it possible to narrow the range of allowed
models with low matter content (the dashed line), while models with typical
values of $\Omega_M$ are still viable for $w_0$ covering almost the full
horizontal axis. On the other hand, the constraint on $\eta_{30}$ is still
too weak so that almost the full region plotted is allowed.

Finally, we consider an hypothetical future SNeIa survey working at the
same photometric accuracy as SNAP and with the same redshift distribution,
but increasing the number of SNeIa up to ${\cal{N}}_{SNeIa} = 6 {\times} 10^4$ as
expected from, e.g., DES \cite{DES}, PanSTARRS \cite{PanSTARRS}, SKYMAPPER
\cite{SKY}, while still larger numbers may potentially be achieved by
ALPACA \cite{ALPACA} and LSST \cite{LSST}. Such a survey can achieve\,:

\begin{displaymath}
(\sigma_1, \sigma_2, \sigma_3, \sigma_4) = (0.02, 0.2, 0.9, 2.7)
\end{displaymath}
so that, with $\sigma_M/\Omega_M \sim 0.1\%$, we get\,:

\begin{displaymath}
(\sigma_{20}, \sigma_{30}) = (0.0015, 0.0016) \ .
\end{displaymath}
Fig.\,\ref{fig: r20} shows that, with such a precision on $\eta_{20}$, the
region of $w_0$ values allowed essentially reduces to the $\Lambda$CDM
value, while, from Fig.\,\ref{fig: r30}, it is clear that the constraint on
$\eta_{30}$ definitively excludes models with low matter content further
reducing the range of $w_0$ values to quite small deviations from the $w_0
= -1$. We can therefore conclude that such a survey will be able to discriminate
between the concordance $\Lambda$CDM model and all the $f(R)$ theories
giving the same cosmographic parameters as quiessence models other than the
$\Lambda$CDM itself.

A similar discussion may be repeated for $f(R)$ models sharing the same
$(q_0, j_0, s_0, l_0)$ values as the CPL model even if it is less intuitive
to grasp the efficacy of the survey being the parameter space multivalued.
For the same reason, we have not explored what is the accuracy on the
double power\,-\,law or HS models, even if this is technically possible.
Actually, one should first estimate the errors on the present day value of
$f(R)$ and its three time derivatives and then propagate them on the model
parameters using the expressions obtained in Sect. VI. The multiparameter
space to be explored makes this exercise quite cumbersome so that we leave
it for a furthcoming work where we will explore in detail how these models
compare to the present and future data.

\section{Conclusions}

The recent amount of good quality data have given a new input to the
observational cosmology. As often in science, new and better data lead to
unexpected discoveries as in the case of the nowadays accepted evidence for
cosmic acceleration. However, a fierce and strong debate is still open on
what this cosmic speed up implies for theoretical cosmology. The equally
impressive amount of different (more or less) viable candidates have also
generated a great confusion so that model independent analyses are welcome.
A possible solution could come from  the cosmography of the universe rather
than assuming {\it ad hoc} solutions of the cosmological Friedmann
equations. Present day and future SNeIa surveys have renewed the interest
in the determination of the cosmographic parameters so that it is worth
investigating how these quantities can constrain cosmological models.

Motivated by this consideration, in the framework of metric formulation of
$f(R)$ gravity, we have here derived the expressions of the present day
values of $f(R)$ and its first three derivatives as function of the matter
density parameter $\Omega_M$, the Hubble constant $H_0$ and the
cosmographic parameters $(q_0, j_0, s_0, l_0)$. Although based on a third
order Taylor expansion of $f(R)$, we have shown that such relations hold
for a quite large class of models so that they are valid tools to look for
viable $f(R)$ models without the need of solving the mathematically
difficult nonlinear fourth order differential field equations.

Notwithstanding the common claim that we live in the era of {\it precision
cosmology}, the constraints on $(q_0, j_0, s_0, l_0)$ are still too weak to
efficiently apply the program we have outlined above. As such, we have
shown how it is possible to establish a link between the popular CPL
parameterization of the dark energy equation of state and the derivatives
of $f(R)$, imposing that they share the same values of the cosmographic
parameters. This analysis has lead to the quite interesting conclusion that
the only $f(R)$ function able to give the same values of $(q_0, j_0, s_0,
l_0)$ as the $\Lambda$CDM model is indeed $f(R) = R + 2 \Lambda$. If future
observations will tell us that the cosmographic parameters are those of the
$\Lambda$CDM model, we can therefore rule out all $f(R)$ theories
satisfying the hypotheses underlying our derivation of Eqs.(\ref{eq:
f0z})\,-\,(\ref{eq: f3z}). Actually, such a result should not be considered
as a no way out for higher order gravity. Indeed, one could still work out
a model with null values of $f''(R_0)$ and $f'''(R_0)$ as required by the
above constraints, but non\,-\,vanishing higher order derivatives. One
could well argue that such a contrived model could be rejected on the basis
of the Occam's razor, but nothing prevents from still taking it into
account if it turns out to be both in agreement with the data and
theoretically well founded.

If new SNeIa surveys will determine the cosmographic parameters with good
accuracy, acceptable constraints on the two dimensionless ratios $\eta_{20}
\propto f''(R_0)/f(R_0)$ and $\eta_{30} \propto f'''(R_0)/f(R_0)$ could be
obtained thus allowing to discriminate among rival $f(R)$ theories. To
investigate whether such a program is feasible, we have pursued a Fisher
matrix based forecasts of the accuracy future SNeIa surveys can achieve on
the cosmographic parameters and hence on $(\eta_{20}, \eta_{30})$. It turns
out that a SNAP\,-\,like survey can start giving interesting (yet still
weak) constraints allowing to reject $f(R)$ models with low matter content,
while a definitive improvement is achievable with future SNeIa survey
observing $\sim 10^4$ objects thus making it possible to discriminate
between $\Lambda$CDM and a large class of fourth order theories. It is
worth stressing, however, that the measurement of $\Omega_M$ should come
out as the result of a model independent probe such as the gas mass
fraction in galaxy clusters which, at present, is still far from the $1\%$
requested precision. On the other hand, one can also rely on the $\Omega_M$
estimate from the CMBR anisotropy and polarization spectra even if this
comes to the price of assuming that the physics at recombination is
strictly described by GR so that one has to limit its attention to $f(R)$
models reducing to $f(R) \propto R$ during that epoch. However, such an
assumption is quite common in many $f(R)$ models available in literature so
that it is not a too restrictive limitation.

A further remark is in order concerning what kind of data can be used to
constrain the cosmographic parameters. The use of the fifth order Taylor
expansion of the scale factor makes it possible to not specify any
underlying physical model thus relying on the minimalist assumption that
the universe is described by the flat Robertson\,-\,Walker metric. While
useful from a theoretical perspective, such a generality puts severe
limitations to the dataset one can use. Actually, we can only resort to
observational tests depending only on the background evolution so that the
range of astrophysical probes reduces to standard candles (such as SNeIa
and possibly Gamma Ray Bursts) and standard rods (such as the angular
size\,-\,redshift relation for compact radiosources). Moreover, pushing the
Hubble diagram to $z \sim 2$ may rise the question of the impact of
gravitational lensing amplification on the apparent magnitude of the
adopted standard candle. The magnification probability distribution
function depends on the growth of perturbations \cite{Holz} so that one
should worry about the underlying physical model in order to estimate
whether this effect biases the estimate of the cosmographic parameters.
However, it has been shown \cite{R06,Goobar} that the gravitational lensing
amplification does not alter significantly the measured distance modulus
for $z \sim 1$ SNeIa. Although such an analysis has been done for GR based
models, we can argue that, whatever is the $f(R)$ model, the growth of
perturbations finally leads to a distribution of structures along the line
of sight that is as similar as possible to the observed one so that the
lensing amplification is approximately the same. We can therefore argue
that the systematic error made by neglecting lensing magnification is lower
than the statistical ones expected by the future SNeIa surveys. On the
other hand, one can also try further reducing this possible bias using the
method of flux averaging \cite{WangFlux} even if, in such a case, our
Fisher matrix calculation should be repeated accordingly. It is also worth
noting that the constraints on the cosmographic parameters may be tigthened
by imposing some physically motivated priors in the parameter space. For
instance, we can impose that the Hubble parameter $H(z)$ stays always
positive over the full range probed by the data or that the transition from
past deceleation to present acceleration takes place over the range probed
by the data (so that we can detect it). Such priors should be included in
the likelihood definition so that the Fisher matrix should be recomputed
which is left for a forthcoming paper.

Although the present day data are still too limited to efficiently
discriminate among rival $f(R)$ models, we are confident that an aggressive
strategy aiming at a very precise determination of the cosmographic
parameters could offer stringent constraints on higher order gravity
without the need of solving the field equations or addressing the
complicated problems related to the growth of perturbations. Almost 80
years after the pioneering distance\,-\,redshift diagram by Hubble, the old
cosmographic approach appears nowadays as a precious observational tool to
investigate the new developments of cosmology.

\acknowledgments We warmly thank R. Lazkoz, R. Molinaro, A. Stabile and A.
Troisi for the interesting discussions and suggestions related to
this paper.

\appendix

\section{Distance formulae}

We derive here some useful relations for distance related quantities as
function of the redshift $z$ and the cosmographic parameters. Using their
definitions in Eqs.(\ref{eq: cosmopar}), it is easy to get for the fifth
order Taylor expansion of the scale factor\,:

\begin{eqnarray}
\frac{a(t)}{a(t_{0})} & = & 1 + H_{0} (t-t_{0}) - \frac{q_{0}}{2}
H_{0}^{2} (t-t_{0})^{2} \nonumber \\ ~ & + & \frac{j_{0}}{3!} H_{0}^{3}
(t-t_{0})^{3} + \frac{s_{0}}{4!} H_{0}^{4} (t-t_{0})^{4}
\nonumber \\ ~ & + & \frac{l_{0}}{5!} H_{0}^{5} (t-t_{0})^{5} +\emph{O}[(t-t_{0})^{6}
\label{eq: series}
\end{eqnarray}
with $t_0$ the present day age of the universe. Note that Eq.(\ref{eq:
series}) is also the fifth order expansion of $(1 + z)^{-1}$, being the
redshift $z$ defined as $z = a(t_0)/a(t) - 1$. The physical distance
travelled by a photon that is emitted at time $t_{*}$ and absorbed at the
current epoch $t_{0}$ is

\begin{displaymath}
D = c \int dt = c (t_{0} - t_{*})
\end{displaymath}
so that inserting $t_{*} = t_{0} - \frac{D}{c}$ into Eq.(\ref{eq: series})
gives us an expression for the redshift as function of $t_0$ and $D/c$,
i.e. $z = z(D)$. Solving with respect to $D$ up to the fifth order in $z$
gives us the desired expansion for $D(z)$ as\,:

\begin{equation}
D(z) = \frac{c z}{H_{0}} \left\{ \mathcal{D}_{z}^{0} +
\mathcal{D}_{z}^{1} \ z + \mathcal{D}_{z}^{2} \ z^{2} +
\mathcal{D}_{z}^{3} \ z^{3} + \mathcal{D}_{z}^{4} \ z^{4} \right\}
\label{eq: seriesdz}
\end{equation}
with\,:

\begin{displaymath}
{\cal{D}}_{z}^{0} = 1 \ , \nonumber
\end{displaymath}

\begin{displaymath}
{\cal{D}}_{z}^{1} = -(1 + q_0/2) \ , \nonumber
\end{displaymath}

\begin{displaymath}
{\cal{D}}_{z}^{2} = 1 + q_{0} + \frac{q_{0}^{2}}{2} - \frac{j_{0}}{6} \ ,
\end{displaymath}

\begin{eqnarray}
{\cal{D}}_{z}^{3} & = & - \left ( 1 + \frac{3}{2}q_{0}+
\frac{3}{2}q_{0}^{2} + \frac{5}{8} q_{0}^{3} \right . \nonumber \\
~ & ~ & - \left . \frac{1}{2} j_{0} -
  \frac{5}{12} q_{0} j_{0} - \frac{s_{0}}{24} \right ) \ , \nonumber
\end{eqnarray}

\begin{eqnarray}
{\cal{D}}_{z}^{4} & = & 1 + 2 q_{0} + 3 q_{0}^{2} + \frac{5}{2} q_{0}^{3} +
\frac{7}{2} q_{0}^{4} \nonumber \\ ~ & - & \frac{5}{3} q_{0}
  j_{0} - \frac{7}{8} q_{0}^{2} j_{0} - j_{0} + \frac{j_{0}^{2}}{12} \nonumber \\
~ & - & \frac{1}{8} q_{0} s_{0}  - \frac{s_{0}}{6} - \frac{l_{0}}{120} \ .
\nonumber
\end{eqnarray}
In typical applications, one is not interested in the physical distance
$D(z)$, but rather in the {\it luminosity distance}\,:

\begin{equation}
D_{L} = \frac{a(t_{0})}{a(t_{0} - D/c)} (a_0 r_{0}) \ ,
\label{eq: dldef}
\end{equation}
or the {\it angular diameter distance}\,:

\begin{equation}
D_{A} = \frac{a(t_{0}- D/c)}{a(t_{0})} \: (a_0 r_{0})
\label{eq: dadef}
\end{equation}
with $a_0 = a(t_0)$ and

\begin{equation}
r_{0}(D) = \left\{
\begin{array}{ll}
\displaystyle{\sin{\int_{t_0 - D/c}^{t_0}{\frac{c dt}{a(t)}}}} & k = 1 \\
~ & ~ \\
\displaystyle{\int_{t_0 - D/c}^{t_0}{\frac{c dt}{a(t)}}} & k = 0 \\
~ & ~ \\
\displaystyle{\sinh{\int_{t_0 - D/c}^{t_0}{\frac{c dt}{a(t)}}}} & k = -1 \\
\end{array}
\right . \ .
\label{eq: rzd}
\end{equation}
Using Eq.(\ref{eq: series}), some cumbersome algebra finally gives\,:

\begin{eqnarray}
\frac{r_{0}(D)}{D/a_0} & = & {\cal{R}}_{D}^{0} +
{\cal{R}}_{D}^{1} \left ( \frac{H_{0} D}{c} \right ) \nonumber \\ ~ & + &
{\cal{R}}_{D}^{2} \left ( \frac{H_{0} D}{c} \right )^{2} +
{\cal{R}}_{D}^{3} \left ( \frac{H_{0} D}{c} \right )^{3} + \nonumber \\ ~ &
+ & {\cal{R}}_{D}^{4} \left ( \frac{H_{0} D}{c} \right )^{4} +
{\cal{R}}_{D}^{5} \left ( \frac{H_{0} D}{c} \right )^{5} \nonumber
\end{eqnarray}
with\,:

\begin{displaymath}
\mathcal{R}_{D}^{0} = 1 \ ,
\end{displaymath}

\begin{displaymath}
\mathcal{R}_{D}^{1} = 1/2 \ ,
\end{displaymath}

\begin{displaymath}
\mathcal{R}_{D}^{2} = \frac{1}{6} \left [ 2 + q_{0}
- \frac{k c^{2}}{H_{0}^{2} a_{0}^{2}} \right ] \ ,
\end{displaymath}

\begin{displaymath}
\mathcal{R}_{D}^{3} = \frac{1}{24} \left [ 6 + 6 q_{0} + j_{0}
- 6 \frac{k c^{2}}{H_{0}^{2} a_{0}^{2}} \right ] \ ,
\end{displaymath}

\begin{displaymath}
\mathcal{R}_{D}^{4}  = \frac{1}{120} \left [ 24 + 36 q_{0} + 6 q_{0}^{2}
+ 8 j_{0} - s_{0} - \frac{5kc^{2}(7 + 2 q_{0})}{a_{0}^{2} H_{0}^{2}}
\right ] \ ,
\end{displaymath}

\begin{eqnarray}
\mathcal{R}_{D}^{5} & = & \frac{24 + 48 q_{0} + 18 q_{0}^{2}
+ 4 q_{0} j_{0} + 12 j_{0} - 2 s_{0} + 24 l_{0}}{144} \nonumber \\ ~ & - &
\frac{3kc^{2}(15 + 10 q_{0} + j_{0})}{144 a_{0}^{2} H_{0}^{2}} \ . \nonumber
\end{eqnarray}
Expressing $D$ into Eq.(\ref{eq: rzd}) as function of $z$ through
Eq.(\ref{eq: seriesdz}) and inserting the result into Eq.(\ref{eq: dldef}),
one obtains the desired fifth order approximation for the Hubble free
luminosity distance $d_L = D_L(z)/(c/H_0)$ as function of the redshift
$z$\,:

\begin{equation}
d_{L}(z) = \mathcal{D}_{L}^{0} z +
\mathcal{D}_{L}^{1} \ z^2 + \mathcal{D}_{L}^{2} \ z^{3} +
\mathcal{D}_{L}^{3} \ z^{4} + \mathcal{D}_{L}^{4} \ z^{5}
\label{eq: dlseries}
\end{equation}
having defined\,:

\begin{displaymath}
\mathcal{D}_{L}^{0} = 1 \ ,
\end{displaymath}

\begin{displaymath}
\mathcal{D}_{L}^{1} = - \frac{1}{2} \left ( -1 + q_{0} \right ) \ ,
\end{displaymath}

\begin{displaymath}
\mathcal{D}_{L}^{2} = - \frac{1}{6} \left(1 - q_{0} - 3q_{0}^{2} + j_{0}
+ \frac{k c^{2}}{H_{0}^{2}a_{0}^{2}}\right) \ ,
\end{displaymath}

\begin{eqnarray}
\mathcal{D}_{L}^{3} & = & \frac{2 - 2 q_{0} - 15 q_{0}^{2} - 15 q_{0}^{3}
+ 5 j_{0} + 10 q_{0} j_{0} + s_{0}}{24} \nonumber \\ ~ & + & \frac{2 k
c^{2} (1 + 3 q_{0})}{24 H_{0}^{2} a_{0}^{2}} \ , \nonumber
\end{eqnarray}

\begin{eqnarray}
\mathcal{D}_{L}^{4} & = & \frac{-6 + 6 q_{0} + 81 q_{0}^{2} + 165 q_{0}^{3} +
105 q_{0}^{4}}{120} \nonumber \\ ~ & + & \frac{10 j_0^{2} - 27 j_{0} - 110
q_{0} j_{0} - 105 q_{0}^{2} j_{0}}{120} \nonumber \\ ~ & - & \frac{15 q_{0}
s_{0} + 11 s_{0} + l_{0}}{120} \nonumber \\ ~ & - & \frac{5kc^{2}(1 + 8
q_{0} + 9 q_{0}^{2} - 2 j_{0})}{120 a_{0}^{2} H_{0}^{2}} \ . \nonumber
\end{eqnarray}
Finally, a similar procedure gives the following approximation for the
Hubble free angular diameter distance $d_A(z) = D_A(z)/(c/H_0)$ to fifth
order in $z$\,:

\begin{equation}
d_{A}(z) = \mathcal{D}_{A}^{0} z +
\mathcal{D}_{A}^{1} \ z^2 + \mathcal{D}_{A}^{2} \ z^{3} +
\mathcal{D}_{A}^{3} \ z^{4} + \mathcal{D}_{A}^{4} \ z^{5}
\label{eq: daseries}
\end{equation}
having set\,:

\begin{displaymath}
\mathcal{D}_{A}^{0} = 1 \ ,
\end{displaymath}

\begin{displaymath}
\mathcal{D}_{A}^{1} = - \frac{1}{2} \left( 3 + q_{0} \right ) \ ,
\end{displaymath}

\begin{displaymath}
\mathcal{D}_{A}^{2} = \frac{1}{6}
\left [ 11 + 7 q_{0} + 3q_{0}^{2} - j_{0} - \frac{k c^{2}}{H_{0}^{2}a_{0}^{2}} \right ] \ ,
\end{displaymath}

\begin{eqnarray}
\mathcal{D}_{A}^{3} & = & - \frac{50 + 46 q_{0} + 39 q_{0}^{2} + 15 q_{0}^{3}
- 13 j_{0} - 10 q_{0} j_{0} - s_{0}}{24} \nonumber \\ & + & \frac{2 k c^{2} (5 + 3 q_{0})}
{24 H_{0}^{2} a_{0}^{2}} \ , \nonumber
\end{eqnarray}

\begin{eqnarray}
\mathcal{D}_{A}^{4} & = & \frac{274 + 326 q_{0} + 411 q_{0}^{2} + 315 q_{0}^{3} + 105 q_{0}^{4}}
{120} \nonumber \\ ~ & + & \frac{10 j_0^{2} - 137 j_{0} - 210 q_{0} j_{0} -
105 q_{0}^{2} j_{0} - 15 q_{0} s_{0} - 21 s_{0} - l_{0}}{120} \nonumber \\
~ & - & \frac{5kc^{2}(17 + 20 q_{0} + 9 q_{0}^{2} - 2 j_{0})}{120 a_{0}^{2}
H_{0}^{2}} \ . \nonumber
\end{eqnarray}
Using such expressions (for $k = 0$ since we have assumed a flat universe
in the text), it is then straightforward to compute the quantities entering
the Fisher matrix so that no numerical integration and differentation are
needed.

\end{document}